\documentclass[aps,prc,twocolumn,amsmath,amssymb,superscriptaddress,showpacs,showkeys,preprintnumbers]{revtex4-2}
\usepackage{gensymb}
\usepackage{graphicx,color}
\usepackage{amssymb}
\usepackage{enumerate}
\usepackage{verbatim}
\usepackage{dsfont}
\usepackage{float}
\usepackage[outdir=./]{epstopdf}
\usepackage{hyperref}
\usepackage{cleveref}
\usepackage{braket}
\usepackage{rotating}
\usepackage{natbib}
\begin{document}
  \newcommand {\nc} {\newcommand}
  \nc {\Sec} [1] {Sec.~\ref{#1}}
  \nc {\IBL} [1] {\textcolor{black}{#1}} 
  \nc {\IR} [1] {\textcolor{red}{#1}} 
  \nc {\IB} [1] {\textcolor{blue}{#1}} 
  \nc {\IG} [1] {\textcolor{green}{#1}}
  
\title{The complete quantification of parametric uncertainties in (d,p) transfer reactions}

\author{M.~Catacora-Rios}
\affiliation{Department of Physics and Astronomy, Michigan State University, East Lansing, MI 48824-1321}
\affiliation{Facility for Rare Isotope Beams, Michigan State University, East Lansing MI, 48824}
\author{A.~E.~Lovell}
\affiliation{Theoretical Division, Los Alamos National Laboratory, Los Alamos, NM 87545}
\author{F.~M.~Nunes}
\affiliation{Department of Physics and Astronomy, Michigan State University, East Lansing, MI 48824-1321}
\affiliation{Facility for Rare Isotope Beams, Michigan State University, East Lansing MI, 48824}

\date{\today}
\preprint{LA-UR-22-32946}


\begin{abstract}
\begin{description}
\item[Background] Deuteron-induced transfer reactions are a popular probe in nuclear structure and nuclear astrophysics studies. The interpretation of these transfer measurements relies on reaction theory that takes as input effective interactions between the nucleons and the target nucleus. 
\item[Purpose] Previous work quantified the uncertainty associated with the optical potentials between the nucleons and the target.  In this study, we extend that work by also including the parameters of the mean field associated with the overlap function of the final bound state, thus obtaining the full parametric uncertainty on transfer observables.
\item[Method] We use Bayesian Markov Chain Monte Carlo simulations to obtain parameter posterior distributions. We use elastic-scattering cross sections to constrain the optical potential parameters and use the asymptotic normalization coefficient of the final state to constrain the bound state interaction.
We then propagate these posteriors to the transfer angular distributions and obtain confidence intervals for this observable. 
\item[Results] We study  (d,p) reactions on $^{14}$C, $^{16}$O, and $^{48}$Ca at energies in the range $E_d=7-24$ MeV. Our results show a strong reduction in uncertainty by using the asymptotic normalization coefficient as a constraint, particularly for those reactions most sensitive to ambiguities in the mean field. For those reactions, the importance of constraining the bound state interaction is equal to that of constrain the optical potentials. The case of $^{14}$C is an outlier because the cross section is less sensitive to the nuclear interior.
\item[Conclusions] When minimal constraints are used on the parameters of the nucleon-target interaction, the $1\sigma$ uncertainties on the differential cross sections are very large ($\sim 140-185$\%). However, if  elastic-scattering data and the asymptotic normalization coefficient are used in the analysis, with an error of $10$\% ($5$\%), this uncertainty reduces to $\sim 30$\% ($\sim 15$\%). 
\end{description}
\end{abstract}


\maketitle

\section{Introduction}
\label{intro}

For decades, transfer reactions have been successfully used as a  probe in nuclear structure and nuclear astrophysics \cite{bardayan2016,Wimmer_2018}. While many of these studies have focused on investigating single-particle states in nuclei (e.g. \cite{wuosmaa2010,walter2019}), recent efforts  \cite{szwec2021,kay2021} have also explored transfer as a probe for significantly-deformed nuclei. In either case, the interpretation of the results rely on a reaction model. 

Theoretical advances for transfer reactions have focused primarily on deuteron-induced reactions (e.g. A(d,p)B, A(d,n)C). 
While there are experiments that still use the distorted-wave Born approximation (DWBA) in the analysis (a perturbative method which, in its standard first-order implementation, simplifies the deuteron incoming wave to the elastic channel \cite{ReactionsBook}), nowadays most studies include deuteron breakup non-perturbatively in the reaction mechanism because it is known to be important. Amongst these non-perturbative methods is the adiabatic wave approximation (ADWA) \cite{Johnson1974}. ADWA treats the excitation energy of the deuteron adiabatically and captures the three-body dynamics in the region where it is necessary.
In ADWA, the inputs are the pairwise interactions: the nucleon-target optical potentials ($U_{nA}$ and $U_{pA}$) and the effective interactions describing the two relevant bound states (the deuteron $V_{np}$ and the final state $V_{nA}$). These input interactions are not well known and introduce large uncertainties.

The need for uncertainty quantification in reactions has been identified as an important priority in the community \cite{whitepaper}. To extract meaningful information from transfer reactions, be it orbital occupancies or capture rates for astrophysics, it is crucial to know the theoretical uncertainties. Over the last few years, significant effort has been devoted to quantifying uncertainties on the nucleon optical potential when using elastic scattering as constraint \cite{lovell2015,lovell2017,lovell2018,king2018,king2019,catacora2019,lovell2020,catacora2021}. The Bayesian approach first explored in the context of nuclear reactions by Lovell et al. \cite{lovell2018} offers a rich set of statistical tools to explore parameter space and provide diagnostics for reducing  uncertainties. Within the Bayesian analysis, propagation of  optical model uncertainties to (d,p) transfer observables is straightforward (e.g. \cite{lovell2018,king2019,catacora2019}). However, so far, the uncertainties coming from the bound state descriptions have not been quantified. 

While the deuteron bound state is comparatively well known, usually the final state being populated through (d,p) or (d,n) is not. Earlier studies have pointed out the large ambiguity associated with the choice of parameters for the mean field in the final state \cite{combined2005,combined2008}. As a consequence, the combined method was suggested as a way to reduce this ambiguity: the analysis of transfer (d,p) should be combined with an independent peripheral measurement from which the asymptotic normalization of the final state (the so-called Asymptotic Normalization Coefficient, ANC) can be extracted. 
Amongst the various peripheral reactions that can be used to extract ANCs are Coulomb dissociation and direct neutron capture reactions (e.g. \cite{c15-cd}), in addition to sub-Coulomb transfer reactions \cite{combined2008,c14dp-anc}.  Using the constraint on the ANC in the analysis of the non-peripheral transfer reaction provides a better handle on the above-mentioned ambiguities.
However, it should be noted that none of these earlier works \cite{combined2005,combined2008} contain a statistical analysis of uncertainties. 

In this work, we perform a Bayesian analysis of (d,p) transfer reactions taking into account both the parametric uncertainties associated with the optical potentials and the mean field describing the final bound state. We constrain the bound state with an independent extraction of the ANC, and we constrain the optical potential with elastic scattering, as done before. We then discuss the relative impact of these two sources of uncertainty in the resulting transfer cross sections. To span a variety of cases, we consider one-neutron transfer reactions on $^{14}$C, $^{16}$O, and $^{48}$Ca at energies in the range $E_d=7-24$ MeV  for which real data exist (e.g. \cite{48Can12,48Cap12,48Cap21,16On7,16Op7,14Cp14,14Cp8}).
In section \ref{TF}, we briefly present the theoretical framework used and, in Section \ref{ND}, we introduce the numerical details of the calculations. Results are presented and discussed in Section \ref{Results}, and the conclusions are drawn in Section \ref{Conclusion}.



\section{Theoretical Framework}
\label{TF}

\subsection{Statistical Considerations}
\label{RT}

\subsubsection{Bayesian Methods}



Unlike frequentist methods, Bayesian statistics give the probability of a single occurrence given a model and some prior information. This provides a robust methodology to study uncertainty propagation, model comparison, and model mixing. Bayes' Theorem states that for some hypothesis $H$, model $M$, and experimental data $D$ \cite{band},
\begin{equation}
    p(H|D,M)=\frac{p(D|H,M)p(H|M)}{p(D|M)}.
\end{equation}
That is, the posterior probability $P(H|D,M)$ of a hypothesis given some data and model is equal to the prior distribution $p(H|M)$ (containing information known about the model before looking at the data) times the likelihood $p(D|H,M)$ (containing information of the goodness of the fit between the hypothesis of the model and the  data). The Bayesian factor $p(D|M)$ in the denominator is the sum of all possible hypotheses of the model space allowed by the prior information and weighted by the likelihood. In previous studies, Bayesian methods have been used to compare uncertainties coming from different data sets and observables \cite{catacora2019,lovell2020}, including the use of the Bayes' factor \cite{catacora2021}. 

In the work presented here, we are concerned with quantifying the uncertainties of transfer reactions using few-body methods. We perform this analysis in a two-step process. First, we use the Bayesian prescription to optimize independently the parameters of our model using several elastic-scattering and ANC data sets. This will allow us to quantize the uncertainties in the nucleon-target optical potentials describing the interactions of our model (Section IIA2 and IIB1). Second, we propagate the quantified uncertainties from the optimization procedure to the different parts of the few-body transfer matrix using ADWA (Section IIB2). 

\subsubsection{Optical Model Optimization}
\label{sec:OMP}



 In this work, we use the optical model to describe the effective interactions of the nucleon-target scattering states. Optical potentials contain real and imaginary parts describing the elastic channel and absorption into non-elastic channels. The nuclear part of the optical potential typically contains three parts:  a volume term, a surface term, and a spin-orbit term, all of Woods-Saxon or derivative of Woods-Saxon shape, each term parametrized by a depth, a radius and a diffuseness:
\begin{eqnarray}\label{WS}
    U(R)&=& -V_o f_{WS}(R;r_o,a_o)+i W_v f_{WS}(R;r_v,a_v) \nonumber \\
    &+& i 4 a_s W_{s} \frac{d}{dR} f_{WS}(R;r_s,a_s) \nonumber \\
    &+& V_{SO} \frac{1}{R} \frac{d}{dR} f_{WS}(R;r_{so},a_{so}) + V_{C}(R,r_{c}),
\end{eqnarray}
where $f_{WS}(R;r_i,a_i)=\frac{1}{1+\exp((r-r_i)/a_i)}$, $V_{SO}$ contains the $(\vec L  \cdot \vec S)$ term and $V_C$ is the point-sphere Coulomb interaction. The optimization using Bayesian methods will be performed on the parameters of the volume and surface terms (first three terms in Eq. \eqref{WS}), while the parameters of the spin-orbit and Coulomb terms are kept fixed \cite{ReactionsBook}.

\subsection{Reaction Theory}
\label{RT}

As mentioned in the previous subsection, we are interested in transfer reactions. In particular, we look at one-neutron transfer reactions of the form $A(d,p)B(g.s.)$, where the final state of  $B=A+n$ is in the ground state and is described as a single-particle state. The low binding energy of the deuteron allows us to describe $(d,p)$ reactions using the three-body Hamiltonian of the system $n+p+A$: 

\begin{equation}
    \mathcal{H}_{3B} = T_R + T_r + U_{nA} + U_{pA} + V_{np} \;,
\end{equation}
where $U_{pA}$ and $U_{nA}$ are the effective interactions of the nucleon-target systems and $V_{np}$ is the nucleon-nucleon interaction. $T_R$ and $T_r$ denote the two-body kinetic operators for the deuteron-target and n-p systems respectively, where $r$ is the relative distance in the n-p system and $R$ is the distance between the center of mass of the n-p system and the target. In the three-body post form \cite{ReactionsBook}, we can write the T-matrix as:
\begin{equation}\label{T_gen}
    T^\mathrm{exact}= \bra{\phi_{I_AI_B}\chi_{pB}^{(-)}}V_{np}+U_{nA}-U_{pB}\ket{\Psi^{exact}}.
\end{equation}
Here, $\Psi^{exact}$ is the full solution of the three-body $A+n+p$ problem in the incident channel, $\phi_{I_AI_B}$ is the overlap of the $B=A+n$ and $A$ bound-state wavefunctions, and $\chi^{(-)}_{pB}$ is the proton distorted wave of the exit channel.



\subsubsection{Asymptotic Normalization Coefficient}

A central element of the analysis of the $A(d,p)B(g.s.)$ reaction, not included in previous uncertainty propagation analyses, is the many-body radial overlap function $\phi_{I_AI_B}$ which depends on the vector-radius $\mathbf{r}$ connecting the center of mass of $A$ and $n$. The wavefunction for the $B=A+n$ system is a bound state, and so it behaves as a spherical Hankel function in the exterior region $(r>R)$ where the nuclear potential is negligible, 
\begin{equation}
    \phi_{I_AI_B(lj)}(r)\stackrel{r>R}{\approx} i\kappa C_{lj} h_{l}(i\kappa r)\;.
\end{equation}
In the above equation, we have $\kappa=\sqrt{2\mu_{An}\epsilon_{An}}$ where $\mu_{An}$ and $\epsilon_{An}$ are the reduced mass and binding energy of the $A+n$ system. We also introduce the quantum numbers $l,j$ for $An$ relative orbital and total angular momentum. Finally, $C_{lj}$ is the ANC. 
Assuming that the many-body overlap function is proportional to the single-particle function, the experimental ANC, $C$, and the spectroscopic factor, $S$, can be related via the following equation:
\begin{equation}
    C^{2}_{lj}=S_{n_r l j}b^{2}_{n_r l j}.
    \label{eq:anc}
\end{equation}
Note that here we used the asymptotic properties of the single-particle wavefunction:
\begin{equation}
    u^{nA}_{n_r,l,j}(r)\stackrel{r>R}{\approx} i\kappa b_{n_r l j} h_l (i\kappa r)
\end{equation}
where the single-particle wavefunction $u^{nA}_{n_r,l,j}$ is generated by adjusting a Woods-Saxon shaped potential with parameters $V_o,r_o,a_o$ to reproduce the correct separation energy of the bound state with the correct quantum numbers $(n_r,l,j)$. Here $n_r$ is the principal quantum number. We have introduced the asymptotic normalization coefficient of $u^{nA}_{n_r,l,j}$ as $b_{n_r l j}$.
The relationship in Eq.\ref{eq:anc} demonstrates that the ANC is directly related to the spectroscopic factor and therefore can be used to reduce the uncertainties in $(d,p)$ reactions.

\subsubsection{ADWA}

The expression for the T-matrix in Eq. \eqref{T_gen} can be further simplified, as shown by Johnson and Tandy, due to the fact that the exact three-body wavefunction $\Psi^\mathrm{exact}$ is only required within the small range of potential $V_{np}$ \cite{Johnson1974}. One can then use the Weinberg basis to expand the exact three-body wavefunction. Neglecting the excitation of the $n-p$ system in the reaction and using only the first term in the expansion, one obtains the adiabatic form of the T-matrix:
\begin{equation}\label{T_ad}
    T = \bra{u_{nA}\chi_{pB}^{(-)}}V_{np}\ket{\phi_{np}\chi_{d}^{ad}}\;,
\end{equation}
where we have dropped the remnant term in the operator $U_{nA}-U_{pB}$. We also note that, in the above, we have replaced the many-body overlap function by the single-particle bound state. The adiabatic function $\chi^{ad}_{d}$ is generated from the effective adiabatic potential 
\begin{equation}\label{V_ad}
    U^{\text{eff}}_{Ap}= -\bra{\phi_o(r)}V_{np}(U_{nA}+U_{pA})\ket{\phi_o(r)},
\end{equation}
where $\phi_o$ is the first Weinberg state. This method is referred as  ADWA (see \cite{Johnson1974} for a detailed discussion).
In this study we only consider local $V_{nA}$ interactions (see \cite{Titus2016} for work on incorporating non-local interactions into the framework).
Eventually, when comparing to transfer data, one multiplies the r.h.s of Eq. \ref{T_ad} by the spectroscopic factor $S$.

As in previous work (e.g. \cite{catacora2019,lovell2020,catacora2021}), will use elastic-scattering data and our Bayesian procedure to constrain the entrance channel effective-potentials of the nucleon-target system in Eq. \eqref{V_ad}. The same procedure will also constrain the proton distorted wave of the exit channel $\chi^{(-)}_{pB}$ in Eq. \eqref{T_ad}. By now including the ANC in the analysis, we can put a parametric constrain on the last piece of the puzzle, the overlap function of the bound state, $u_{An}$. In this way, we obtain a full parametric uncertainty analysis of $(d,p)$ transfer reactions using Bayesian statistics.


\section{Methods and Numerical Details}
\label{ND}


In our Bayesian analysis, we aim to fit a theoretical model prediction, $\sigma^{\text{th}}(\mathbf{x})$ (corresponding to some set of parameters $\textbf{x}$), to a set of experimental data, $\sigma_{i}^{\text{exp}}$, with some experimental error $\varepsilon_i$. 
The corresponding set of parameters $\mathbf{x}$ for each nucleon-target interaction will be optimized separately using data for each reaction. Two types of reaction data are used in this paper, the ANC and differential elastic-scattering cross sections.


\subsection{ANC Fitting}

 The first data $\sigma^{\text{exp}}_i$ considered here is the ANC extracted from previous work (references  provided in Table II). These ANCs will be used to constrain the single-particle bound state wavefunction in the T-matrix of Eq. \eqref{T_ad}. For this type of data, our theoretical model $\sigma^{\text{th}}$ consists of the single-particle potential for $B=A+n$ described by a real volume term and a spin-orbit term. Our Bayesian analysis of the ANC will only sample the real radius, $r_o$, and real diffuseness, $a_o$, of the volume term. For each two-dimensional parameter draw $\mathbf{x}=(r_o,a_0)$, the real volume depth $V_o$ is adjusted to reproduce the correct neutron separation energy of the system. The spin-orbit term, although included in the model, remains fixed and is thus not part of the Bayesian analysis. The quantum numbers, neutron separation energies and experimental $C^2$s are summarized in Table \ref{tab:ANCinfo}. 
 
 \begin{table}[H]
  \centering
\begin{tabular}{c|c|c|c|c|c}
      Bound State & $s$ & $l$ & $j$ & B.E.(MeV) & $C^{2}_{\text{exp}}$ (fm$^{-1}$)\\
     \hline \hline  $^{15}$C(g.s.) & $0.5$ & $0$ &$0.5$ & $1.280$ & $1.48$\\
     \hline  $^{17}$O(g.s.) & $0.5$ & $2.0$ & $2.5$ & $4.143$ & $0.67$ \\
     \hline $^{49}$Ca(g.s.) & $0.5$  & $1.0$ & $1.5$ & $5.146$ & $32.1$\\
     
\end{tabular}
\caption{Quantum numbers, neutron separation energies, and experimental ANC squared for the cases considered in this Bayesian analysis.}
\label{tab:ANCinfo}
\end{table}

 \subsection{Elastic Fitting}
 
 The second type of data $\sigma^{\text{exp}}_{i}$ used in this work is elastic scattering angular distributions $(d\sigma/d\Omega)$ which constrain the nucleon-target interactions for the incoming and outgoing channels. As in our previous work (see \cite{lovell2018,lovell2020} for details), our optical model $\sigma^{\text{th}}$ will now consist of real and imaginary volume terms, a surface term, a spin orbit term and a Coulomb term as in Eq.(3). We will use elastic mock data as was done in \cite{catacora2019,lovell2020,catacora2021}. The mock data are generated from the Koning-Delaroche (KD) optical model \cite{kd2003}. To fit the mock data, we will optimize parameter sets consisting of the real and imaginary volume terms and the surface term, for a total of 9 parameters, $\mathbf{x}=(V_o,r_o,a_o,W_v,r_w,a_w,W_s,r_s,a_s)$. The spin-orbit  and Coulomb terms are kept fixed to the KD value used to generate the mock data.  Table \ref{tab:dataOverview} summarizes the number of parameters being fitted for each reaction (dim, column three) and the type of data used (column four). 

\begin{table}[H]
 \centering
\begin{tabular}{c|c|c|c}
      Reaction/Bound State & Energy(MeV)  & dim($\mathbf{x}$) & Data Type    \\
     \hline \hline  $^{15}$C(g.s.) & -1.280 & 2 & $C^2_{\text{exp}}$\cite{Goss1975} \\
     \hline  $^{14}$C(p,p)$^{14}$C & 8.5   & 9 & $(d\sigma/d\Omega)_{mock}$ \\
     \hline  $^{14}$C(n,n)$^{14}$C & 8.5  & 9 & $(d\sigma/d\Omega)_{mock}$  \\
     \hline  $^{14}$C(p,p)$^{14}$C & 17  & 9 & $(d\sigma/d\Omega)_{mock}$  \\
     \hline  $^{17}$O(g.s.) & -4.143  & 2 & $C^2_{\text{exp}}$\cite{Keller1961}  \\
     \hline  $^{16}$O(p,p)$^{16}$O & 7.5   & 9 & $(d\sigma/d\Omega)_{mock}$ \\
     \hline  $^{16}$O(n,n)$^{16}$O & 7.5  & 9 & $(d\sigma/d\Omega)_{mock}$  \\
     \hline  $^{16}$O(p,p)$^{16}$O & 15  & 9 & $(d\sigma/d\Omega)_{mock}$ \\
     \hline  $^{49}$Ca(g.s.) & -5.146  & 2 & $C^2_{\text{exp}}$\cite{combined2008} \\
     \hline  $^{48}$Ca(p,p)$^{48}$Ca & 12   & 9 & $(d\sigma/d\Omega)_{mock}$ \\
     \hline  $^{48}$Ca(n,n)$^{48}$Ca & 12  & 9 & $(d\sigma/d\Omega)_{mock}$  \\
     \hline  $^{48}$Ca(p,p)$^{48}$Ca & 24   & 9 & $(d\sigma/d\Omega)_{mock}$ \\
     
\end{tabular}
\caption{Data type (column 4) and number of parameters being optimized independently (column 3) for each of the reactions or bound states (column 1) at the given reaction energy or neutron binding energy (column 2) considered in the Bayesian analysis.}
\label{tab:dataOverview}
\end{table}


\subsection{Numerical Details}

As in previous work, we will use the Metropolis-Hastings Markov Chain Monte Carlo (MH-MCMC, see \cite{lovell2018} for numerical details). Our optical potential parameters are initialized with the Becchetti-Greenlees (BG) parametrization \cite{bg69}. We introduce the following shorthand notation for the percent error in the data, $\varepsilon_{5}=0.05\sigma^\mathrm{exp}$, $\varepsilon_{10}=0.1\sigma^\mathrm{exp}$, and $\varepsilon_{100}=1.0\sigma^\mathrm{exp}$, corresponding to 5$\%$, 10$\%$ and 100$\%$ error on the data.  Our prior distributions are Gaussian and centered around the BG value with a width equal to $20\%$  of the mean distribution. As was done in \cite{catacora2021}, if any of the imaginary potential depths is equal to zero, we set it to 1 MeV and give it a width of 10 MeV. Using the MH-MCMC, we will draw  parameter sets $\mathbf{x}$ until $16000$ sets are accepted and we keep one of every ten of these, these $1600$ pulls will constitute our posterior distributions. The computations are performed using the \texttt{QUILTR} code \cite{lovell2020,fresco}. Using the posterior distributions from the bound state and elastic scattering sampling, we will generate predictions for the ADWA transfer cross section using \texttt{QUILTR} which embeds the transfer code NLAT \cite{nlat}.

\section{Results}
\label{Results}
The main goal of this work is to include the parametric uncertainties of the bound state, on top of the uncertainties arising from the optical model, into the analysis of transfer reactions, thus obtaining the full parametric uncertainty quantification for the predicted observables.

\subsection{Constraining the bound state}
\begin{figure}[p]
    \includegraphics[width=7.5cm]{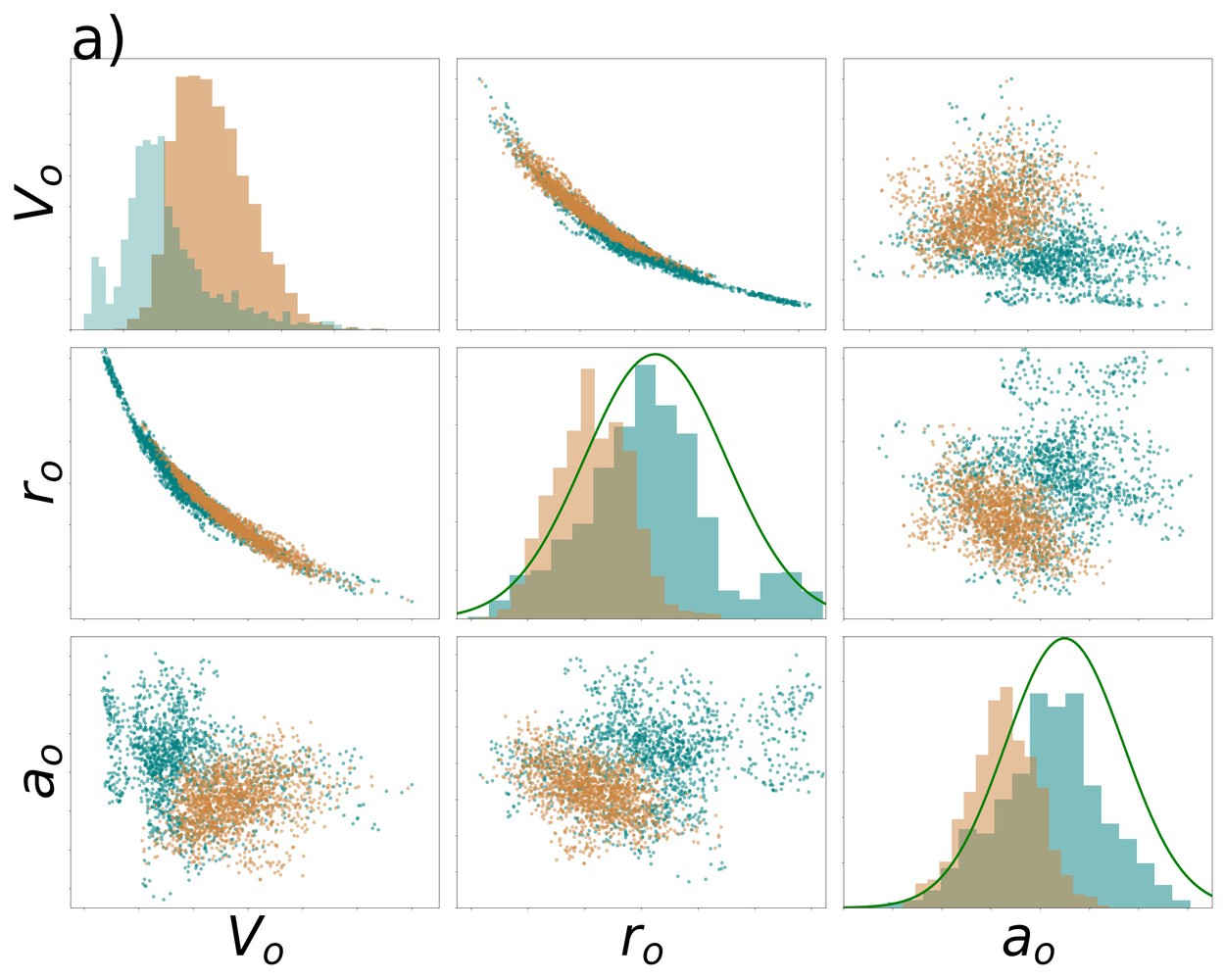}
    \includegraphics[width=7.5cm]
    {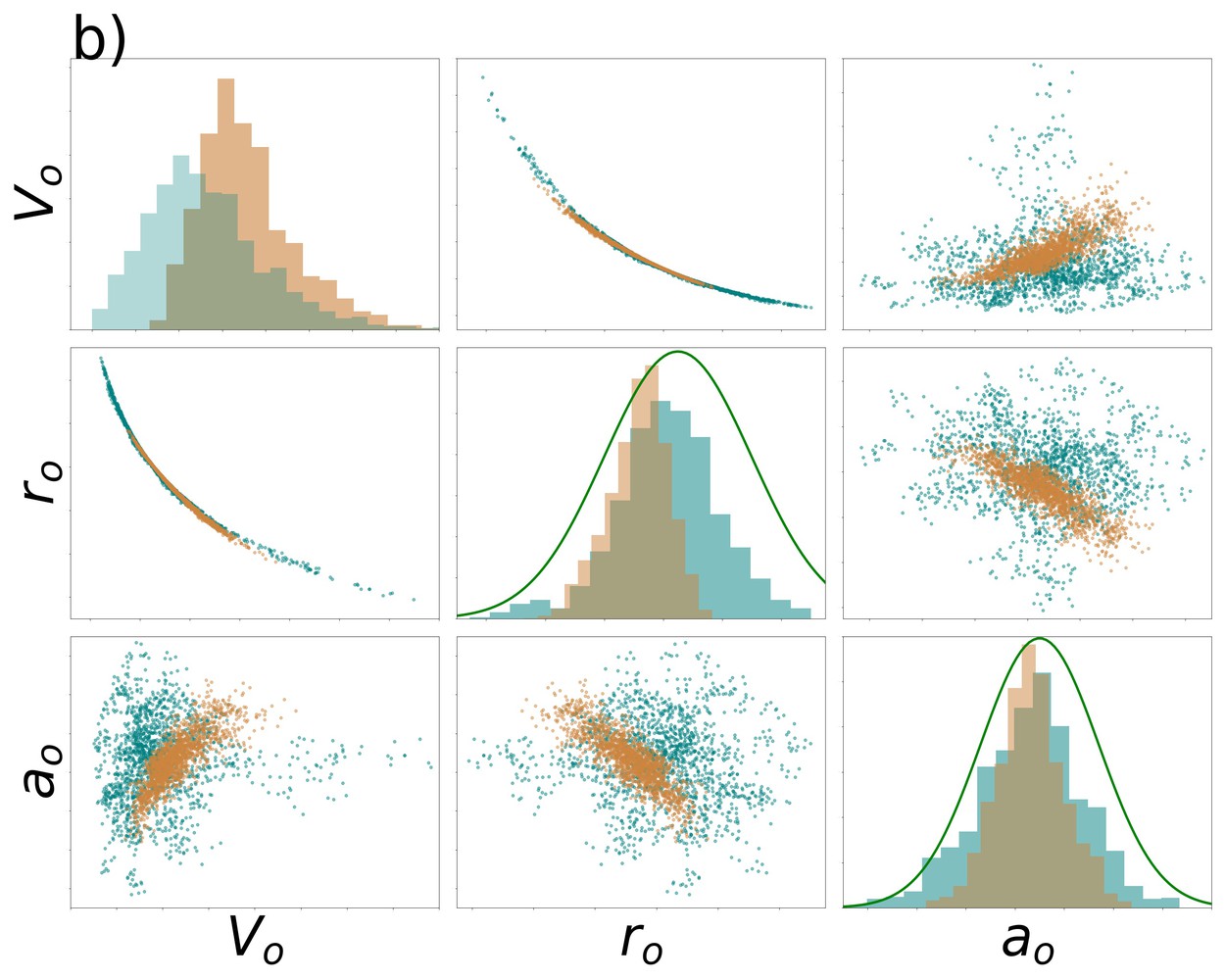}
    \includegraphics[width=7.5cm]{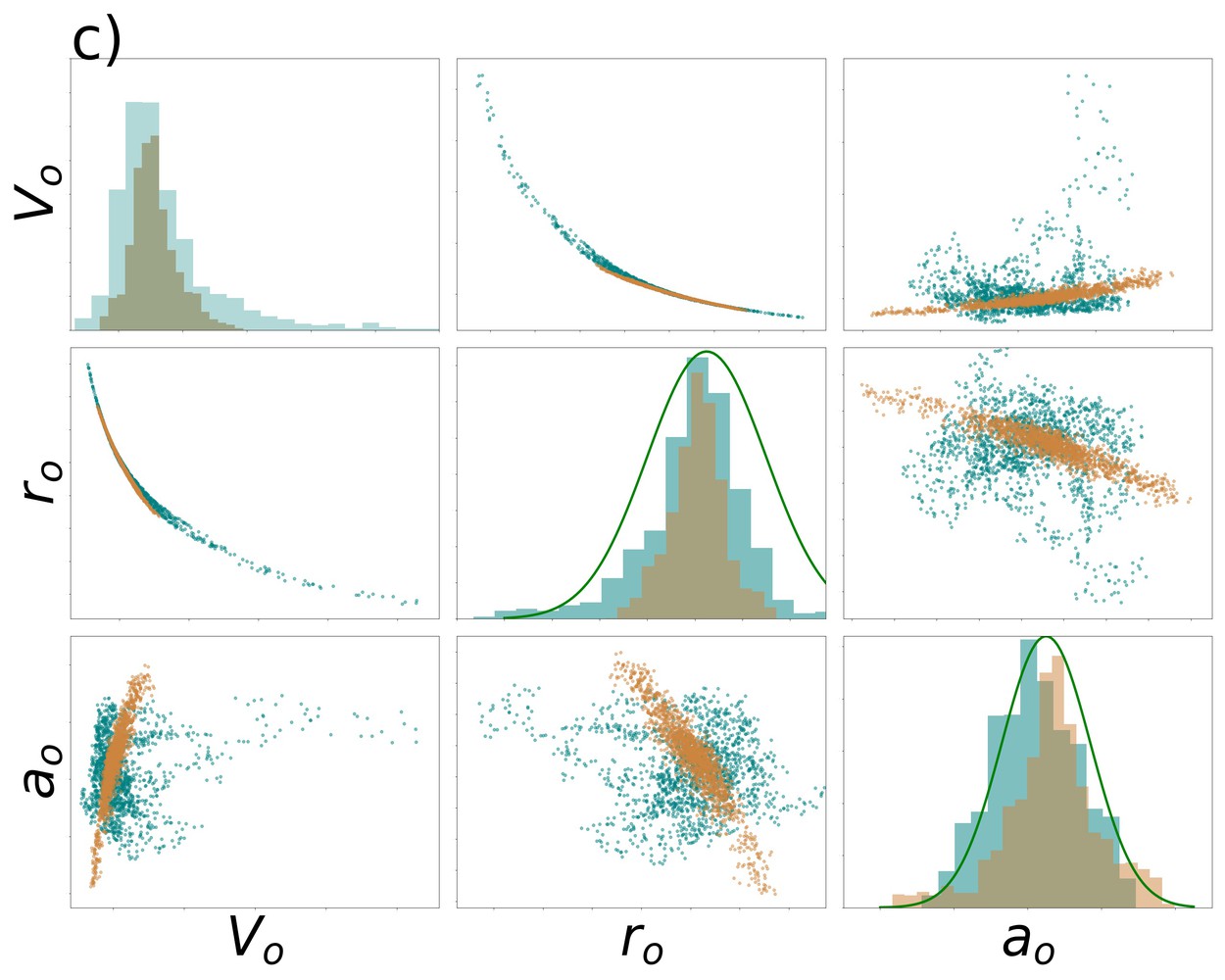}
    \caption{Parameter posteriors and correlations for single-particle parameters of (a) $^{15}$C bound state, (b) $^{17}$O bound state, and (c) $^{49}$Ca bound state, using Bayesian optimization to compare two different experimental errors on $C^2$, $\varepsilon_{10}(\varepsilon=10\%$) in the brown and $\varepsilon_{100}(\varepsilon=100\%)$ in the teal. Histograms on the diagonal correspond to the posterior distributions of the given parameters, the priors are shown in green.}
    \label{fig:correlations}
\end{figure}

As a first step, we quantify the parametric uncertainties obtained by constraining only the bound state using the ANC ($C^2$) and then propagating this uncertainty via the T-matrix in Eq. \eqref{T_ad}. As mentioned in the previous section, the effective interactions for the $A+n$ bound states consist of a real volume term and a spin-orbit term, of these we sample independently  the radius ($r_o$) and diffuseness ($a_o$) adjusting the depth ($V_o$) to reproduce the correct binding energies of the bound state. In Fig. \ref{fig:correlations}, we show the accepted parameter correlations (off-diagonal  plots), posterior distributions (diagonal histograms) and prior distributions (green line in the diagonal histograms). 

We have studied the parameter constrain obtained with the ANC ($C^2$) when the experimental error on $C^2$ is changed from $10\%$ ($\varepsilon_{10}$) in  brown to $100\%$ ($\varepsilon_{100}$) in teal. In panel a), we show the parameters of the $^{15}$C bound state, in panel b), we show the parameters for the $^{17}$O bound state, and in panel c), we show the parameters for the $^{49}$Ca bound state. 

We see that results with a $10\%$ experimental error are shifted for $^{15}$C and $^{16}$O when compared to the posteriors obtained with the $100\%$ experimental error (diagonal plots in Fig. \ref{fig:correlations}). These shifts are quantified in Table III, where for each bound state and $C^2$ experimental error ($\varepsilon_{10}$ or $\varepsilon_{100}$), the mean value and standard deviation of the posteriors are given. As expected, there is also an increase in the width of the posteriors when increasing the experimental error since then a larger set of parameter combinations can be accepted into the Bayesian optimization. The tightened constrain on the parameters from a smaller error on $C^2$ results in a stronger correlation between $r_o$ and $a_o$ (see off-diagonal plots in Fig. 1). One should keep in mind that, because the  depth ($V_o$) is not directly sampled, it is strongly  correlated with $r_o$.

\begin{table}[b]
 \centering
\begin{tabular}{c c c c}
     Bound State($\varepsilon_{C^2}$) & $\mathbf{V_o}(\mathbf{\bar{\sigma}}) $ (MeV)  & $\mathbf{r_o}(\mathbf{\bar{\sigma}})$ (fm) & $\mathbf{a_o}(\mathbf{\bar{\sigma}}) $ (fm) \\
     \hline \hline  $^{15}$C ($\varepsilon_{10}$) & $74.2(13.9)$ &$1.03(0.13)$ &
     $0.52(0.07)$ \\
      $^{15}$C ($\varepsilon_{100}$) & $56.6(18.2)$ &$1.22(0.23)$ &
     $0.62(0.11)$ \\
     \hline   $^{17}$O ($\varepsilon_{10}$)& $65.9(11.0)$ &$1.11(0.10)$ & $0.63(0.07)$ \\
        $^{17}$O ($\varepsilon_{100}$)& $58.9(19.7)$ &$1.21(0.18)$ & $0.63(0.11)$\\
     \hline $^{49}$Ca ($\varepsilon_{10}$) & $51.2(7.23)$ &$1.21(0.12)$ & $0.66(0.13)$ \\
      $^{49}$Ca ($\varepsilon_{100}$) & $57.0(26.4)$ &$1.19(0.21)$ & $0.62(0.12)$ \\
\end{tabular}
\caption{Posterior means and standard deviations optimized with different experimental errors $\varepsilon_{C^2}$ for each bound state studied (column 1), the mean of the real volume term parameters; depth, radius and diffuseness are listed in the second, third and fourth columns respectively (in parenthesis their corresponding standard deviations $\mathbf{\bar{\sigma}}$).   }
\end{table}

Now, we propagate the uncertainties to the differential transfer cross section. First, we will only allow the bound state parameters to vary, while all other interactions are held constant,   including all optical potentials. The 68\% confidence intervals for the predicted transfer cross section are shown in Fig. 2 (left panels): panel a) for $^{14}$C$(d,p)^{15}$C(g.s.), panel  c) for $^{16}$O$(d,p)^{17}$O(g.s.), and panel e) for $^{48}$Ca$(d,p)^{49}$Ca(g.s.).
The right panels b), d) and f) provide the corresponding percent error plots, obtained from the width of the $68\%$ confidence interval divided by the mean at every angle ($\Delta\sigma/\bar{\sigma}$). 
The results using $10$\% error ($100$\%) on the ANC squared are in brown (teal).   We can clearly see that reducing the experimental error from $100\%$ to $10\%$ leads to a large decrease in the uncertainty of the transfer cross section, for all reactions shown. At forward angles, this gain is around a factor of two for $^{14}$C, a factor of four for  $^{16}$O, and even larger for $^{48}$Ca. The $^{48}$Ca reaction is less peripheral than the others, 
and therefore the sensitivity to the details of the single-particle interaction is largest. We have verified that the uncertainty due to the bound state interaction on the transfer cross section scales directly with the error on the $C^2$ for sub-Coulomb reactions, as then the cross section is directly proportional to $C^2$.

\begin{figure}[H]
    \includegraphics[width=4.25cm]{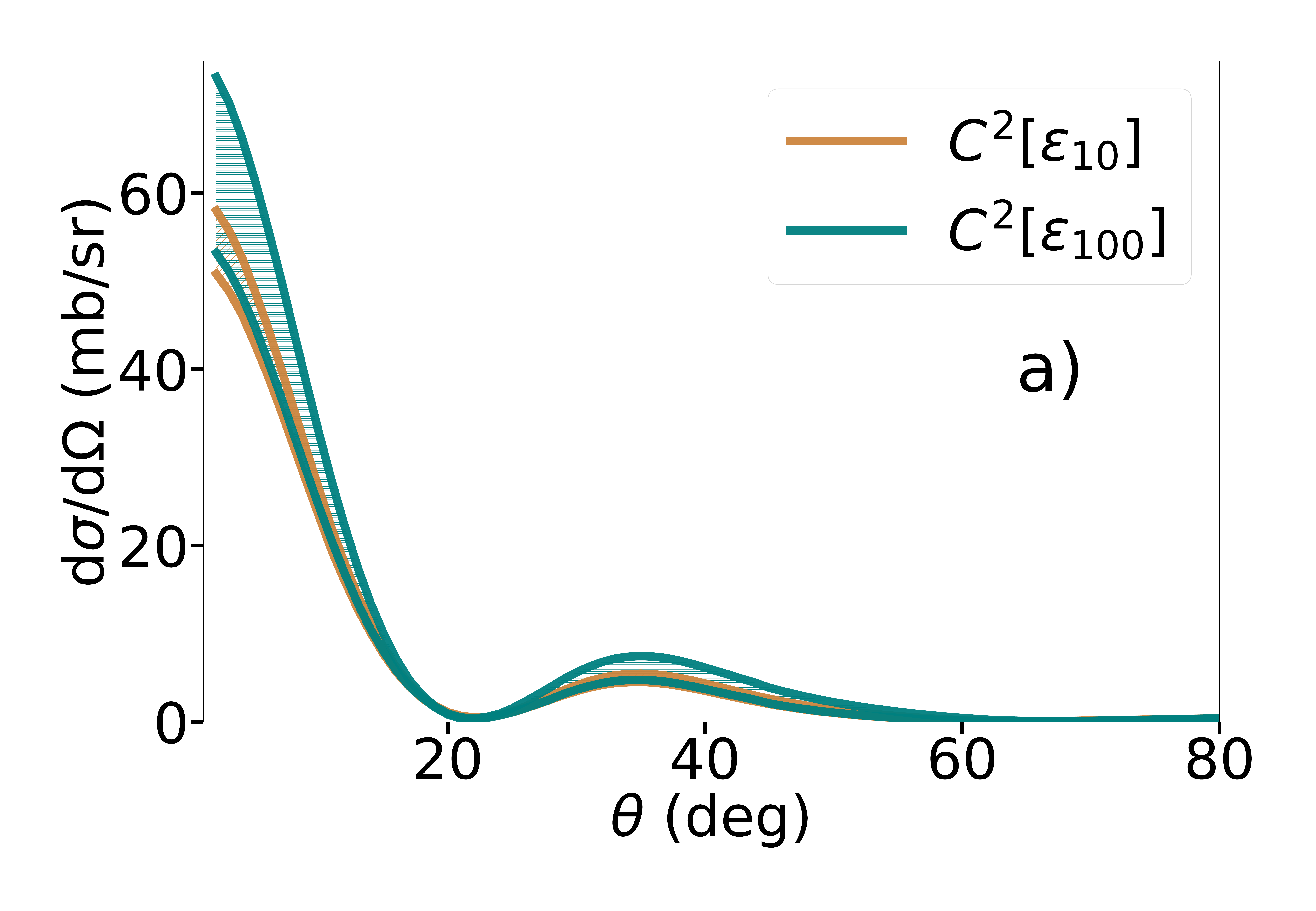}
    \includegraphics[width=4.25cm]{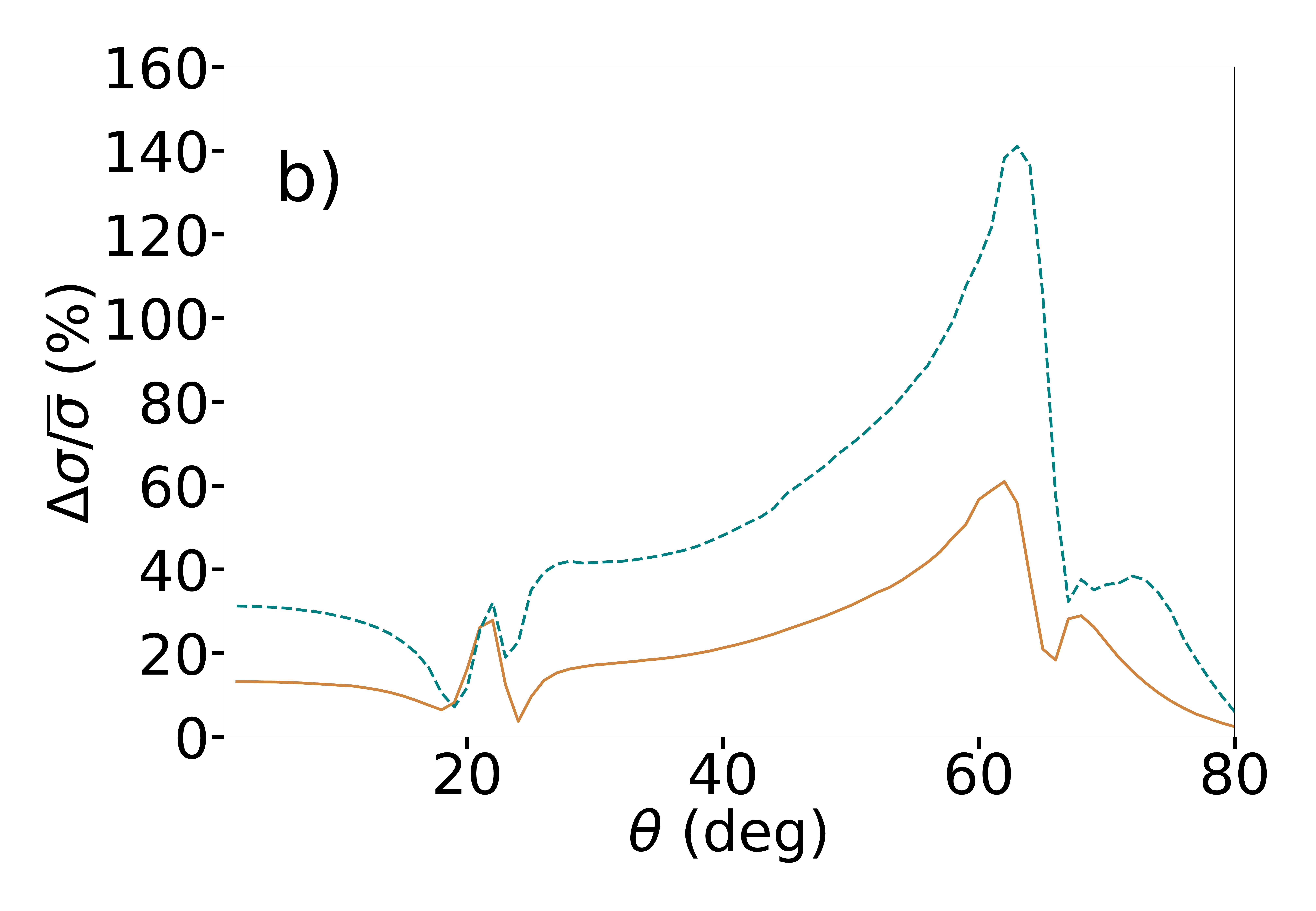}
    \includegraphics[width=4.25cm]{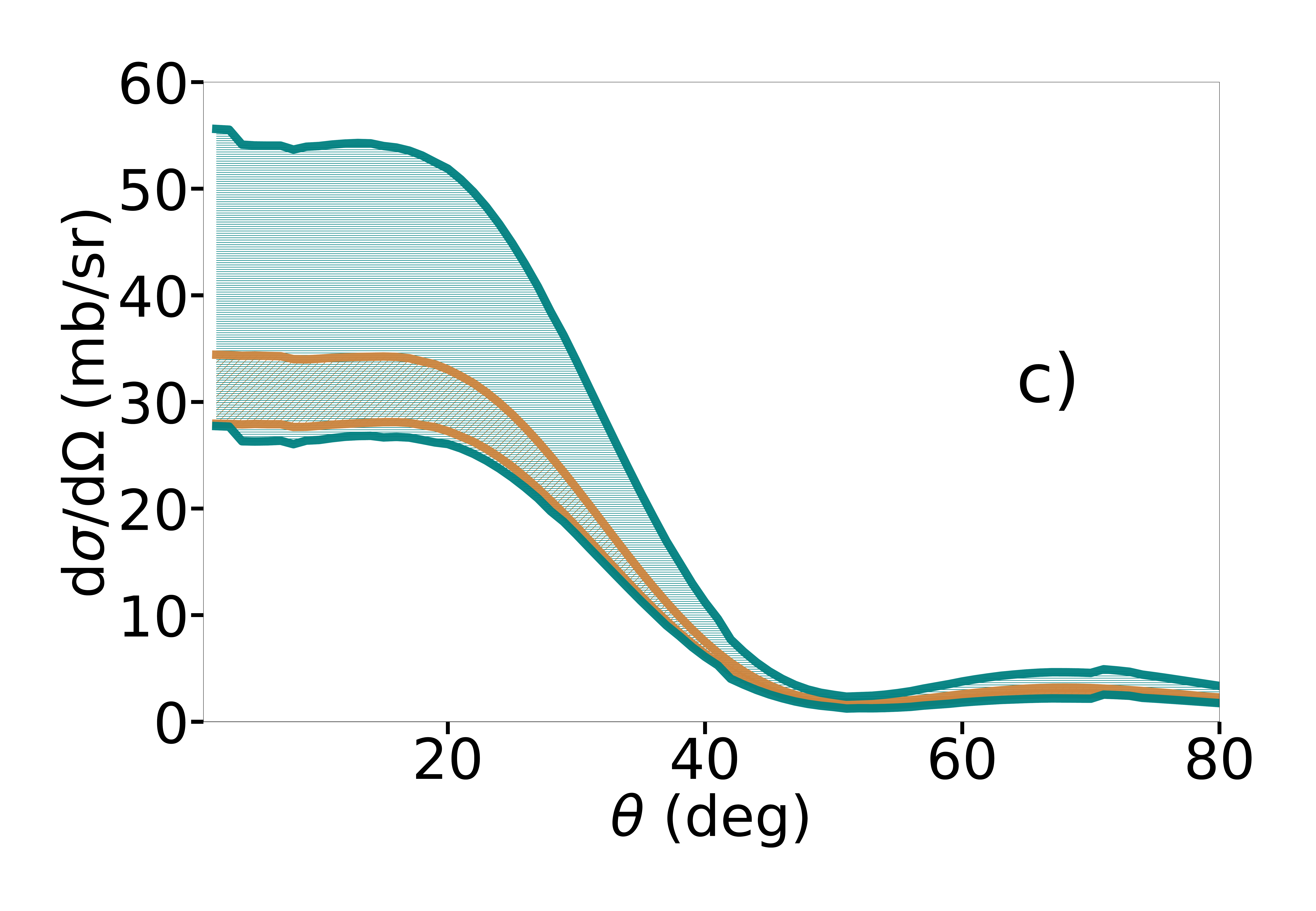}
    \includegraphics[width=4.25cm]{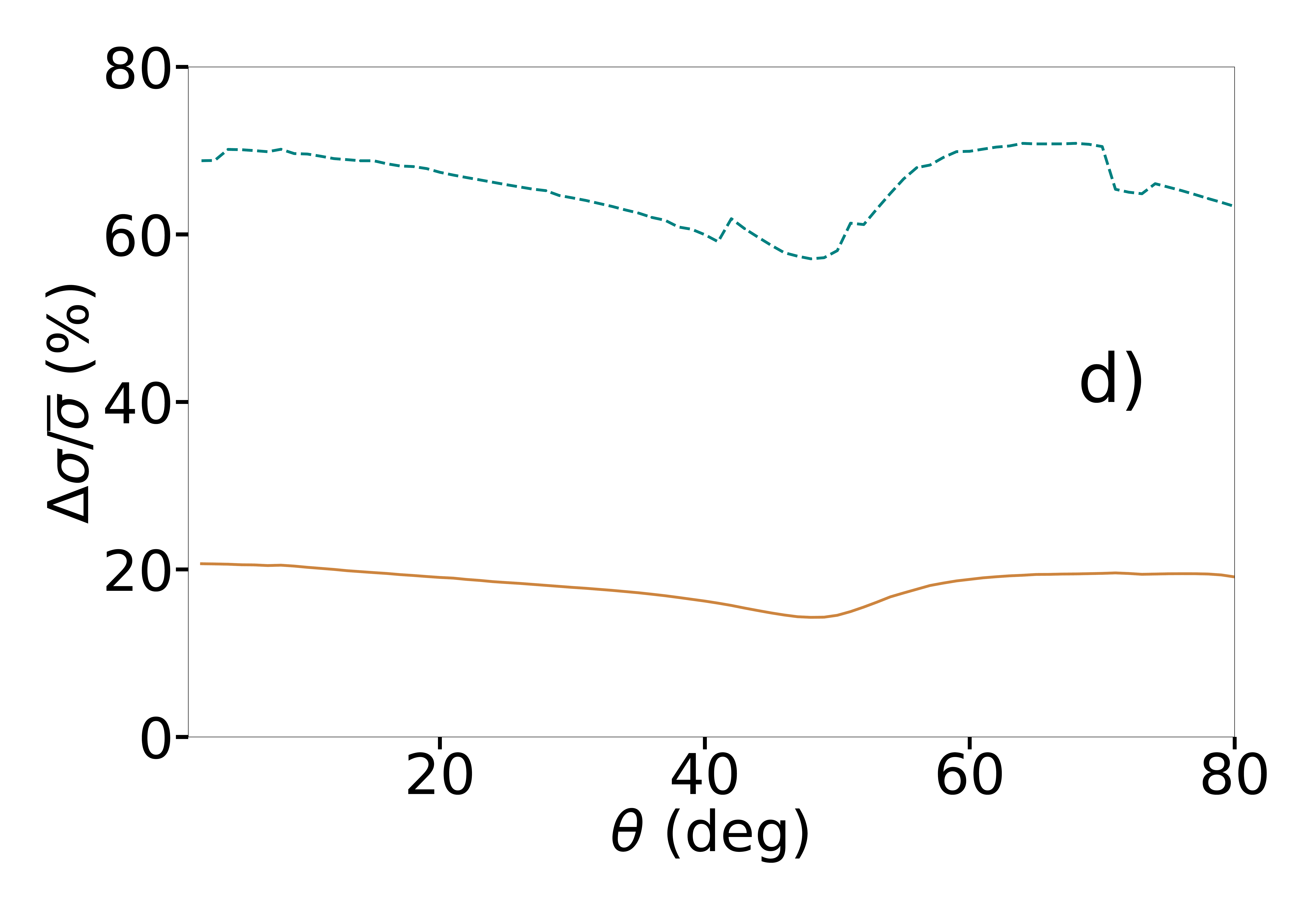}
    \includegraphics[width=4.25cm]{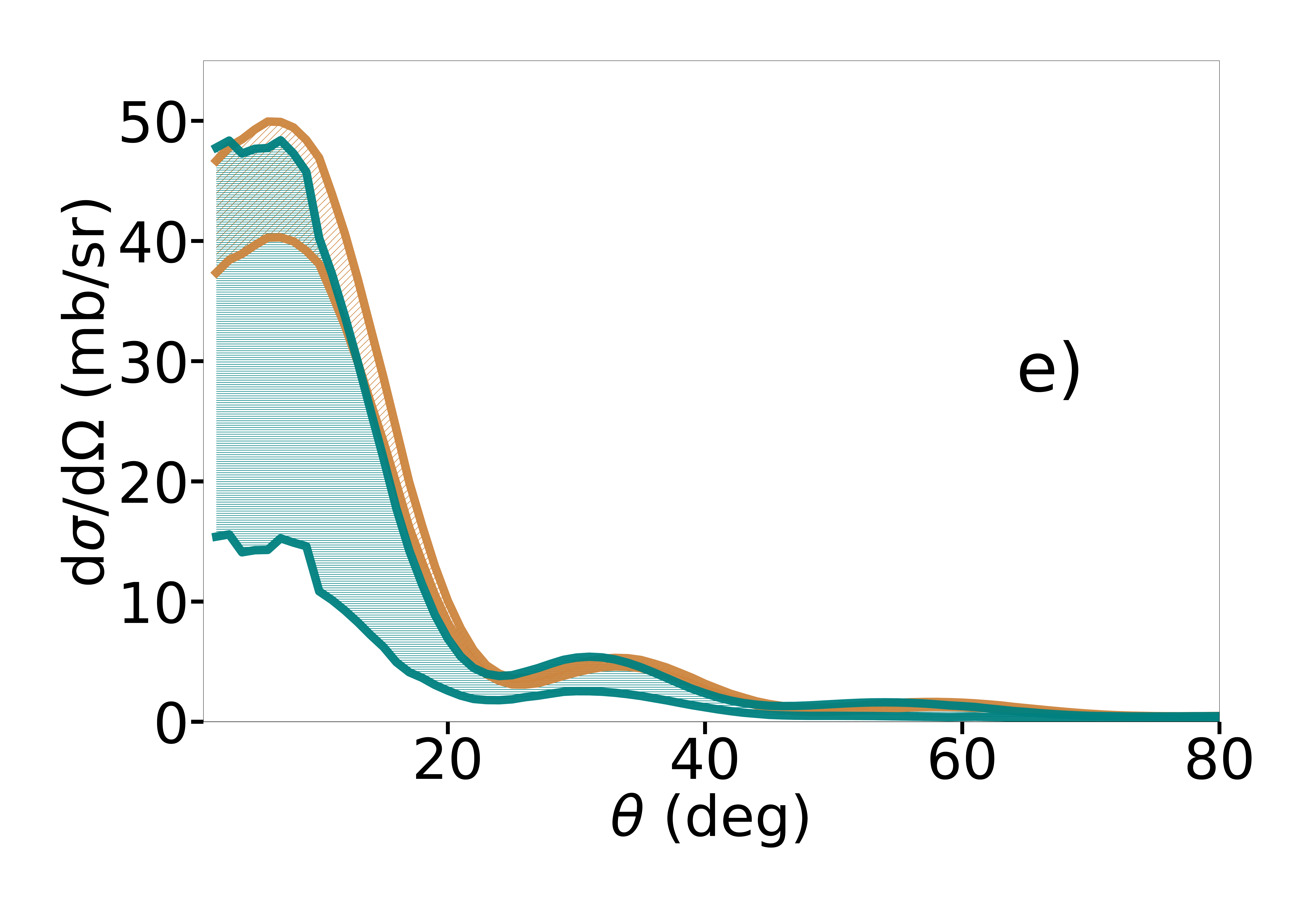}
    \includegraphics[width=4.25cm]{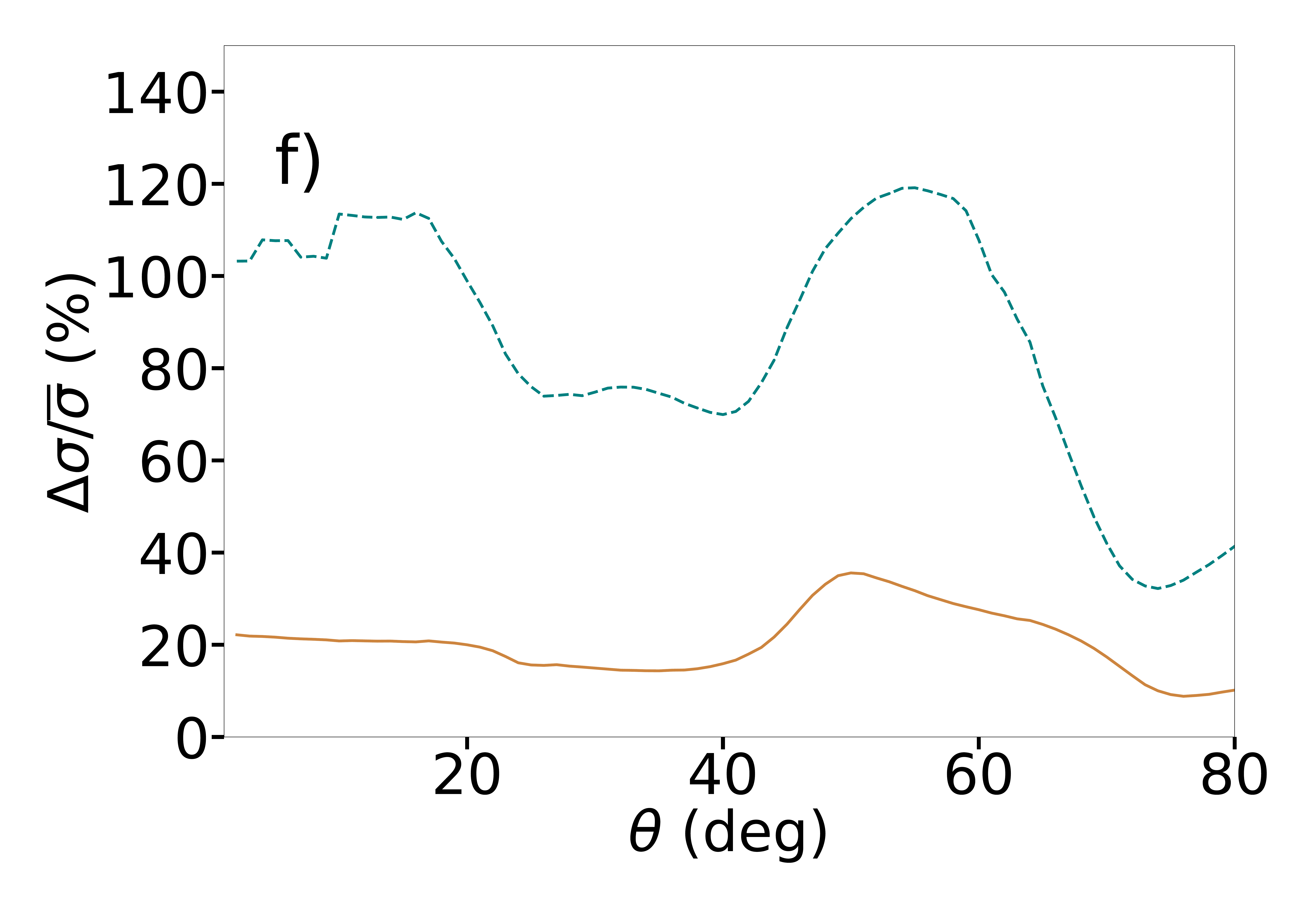}
    \caption{A comparison of results for transfer assuming two $C^2$ experimental data errors, $10\%$ error in brown ($\varepsilon_{10}$) and $100\%$ error in teal ($\varepsilon_{100}$); $^{14}$C(d,p) at 17 MeV 68$\%$ confidence intervals (a) and percentage uncertainty (b); $^{16}$O(d,p) at 15 MeV 68$\%$ confidence intervals (c) and percentage uncertainty (d); and $^{48}$Ca(d,p) at 24 MeV 68$\%$ confidence intervals (e) and percentage uncertainty (f). }
    \label{fig:anc_error}
\end{figure}

\subsection{Elastic vs. ANC constrain}

All results in Sec. IVA assumed the optical potentials are fixed. We have studied optical potential uncertainties before (e.g. \cite{lovell2018,king2019,lovell2020,catacora2021}) so, 
next, we proceed by combining the uncertainties from the bound state interactions with those from the optical potentials, to obtain the full uncertainty quantification (see Table II for a description of the reactions and bound states considered). 

Fig 3. contains transfer results for several assumptions. As before,  $68\%$ confidence intervals are shown on the left panels and the percent uncertainty are shown in the right, $^{14}$C$(d,p)^{15}$C(g.s.) (a+b), $^{16}$O$(d,p)^{17}$O(g.s.) (c+d), and $^{48}$Ca$(d,p)^{49}$Ca(g.s.) (e+f). 
The results in brown assume no uncertainty on the optical potentials constrained by elastic data (these parameters are fixed to the KD values used to generate the data) and only the uncertainties on the bound state parameters are constrained through the $C^2$ taking an error of $10$\% ($\varepsilon_{10}$) just as presented by the brown bands in Fig. 2.
The results in bright green are the reverse, we assume no uncertainty on the bound state interaction and quantify uncertainties only from all optical potentials taking elastic data (el) with  $10$\% uncertainty ($\varepsilon_{10}$). The results in blue incorporate the  uncertainties on both the bound state interaction and the optical potentials assuming the errors on the ANC and elastic scattering data are $10$\% ($C^2+$el). Finally, the results in  gray are the same as the bright-green except that the errors are now $100$\%. Note that the grey band in Fig. 3 does not correspond to the results in teal in Fig. 2. All results in Fig. 2 assume fixed optical potential parameters.
The results in gray in Fig.3 are a lower limit to the "unconstrained" case; they represents only minimal constrain on the parameters of both the bound state and the entrance and exit distorted waves.

From the width of the confidence bands and the percent error plots shown in Fig. 3, we can see that the parametric uncertainties coming from the ANC ($C^2$) and the elastic (el) constraints are of similar order of magnitude and therefore of equal importance. Furthermore, as shown by the wider blue band in Fig. 3, choosing to only propagate the uncertainty of the bound state interaction or the optical potential parameters can lead to an under-representation of the uncertainty in the predicted transfer cross section. Finally, assuming  minimal knowledge constraint on the parameters leads to very large errors. 

These results are summarized in Table IV, which provides the  percent error of the predicted transfer cross section at the first peak of the angular distribution for all cases considered. We choose to quantify the error at this specific angle because the spectroscopic factor ($S$) is typically extracted around this point:  the reaction and the beam energy  are given in column 1 and 2 respectively, the data and experimental error used to constrain the parametric uncertainties is provided in column 3, and columns 4 and 5 show the percent error at the peak for $68\%$ and $95\%$ confidence intervals. 

\begin{figure*}[t]
    \includegraphics[width=8.5cm]{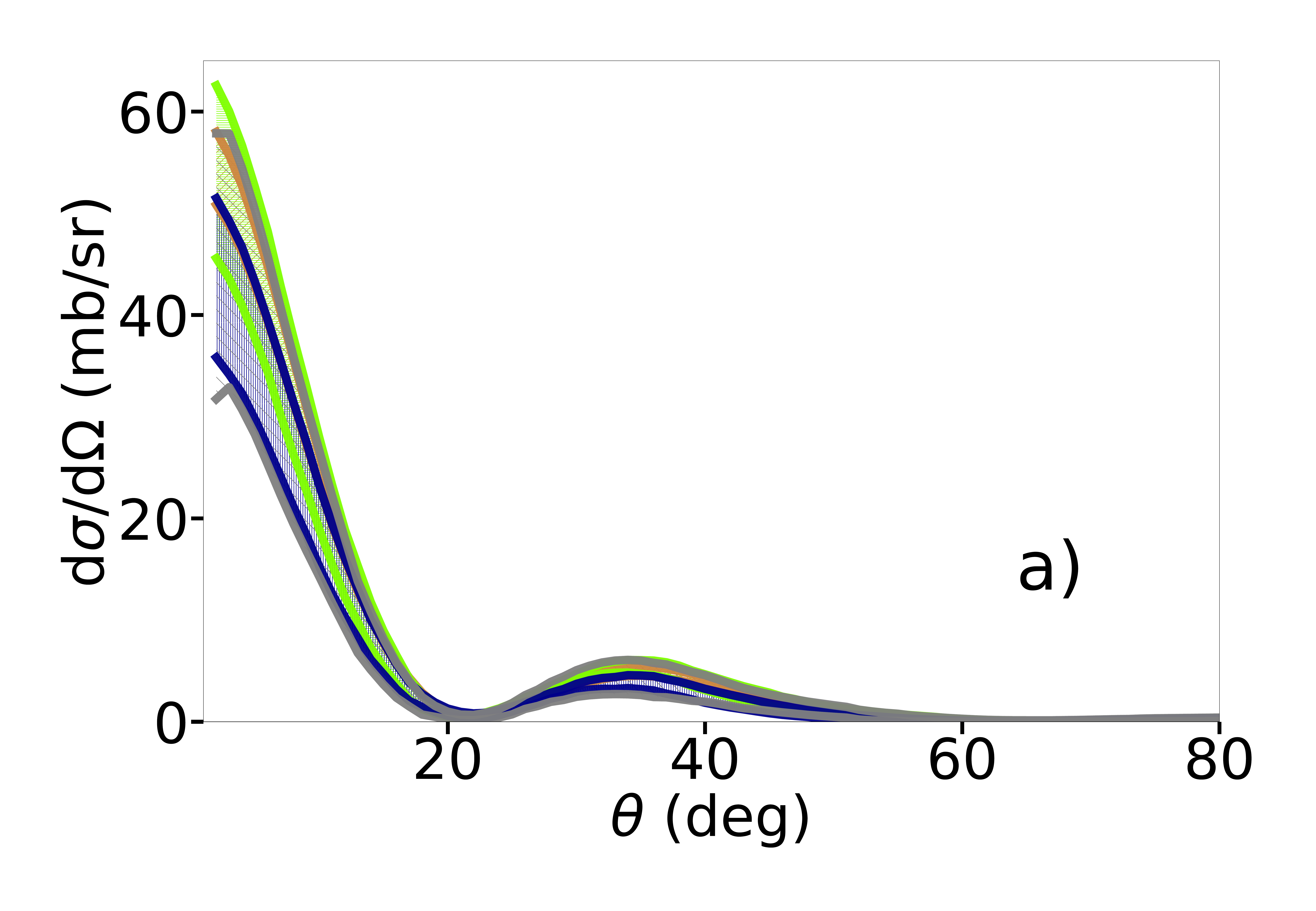}
    \includegraphics[width=8.5cm]{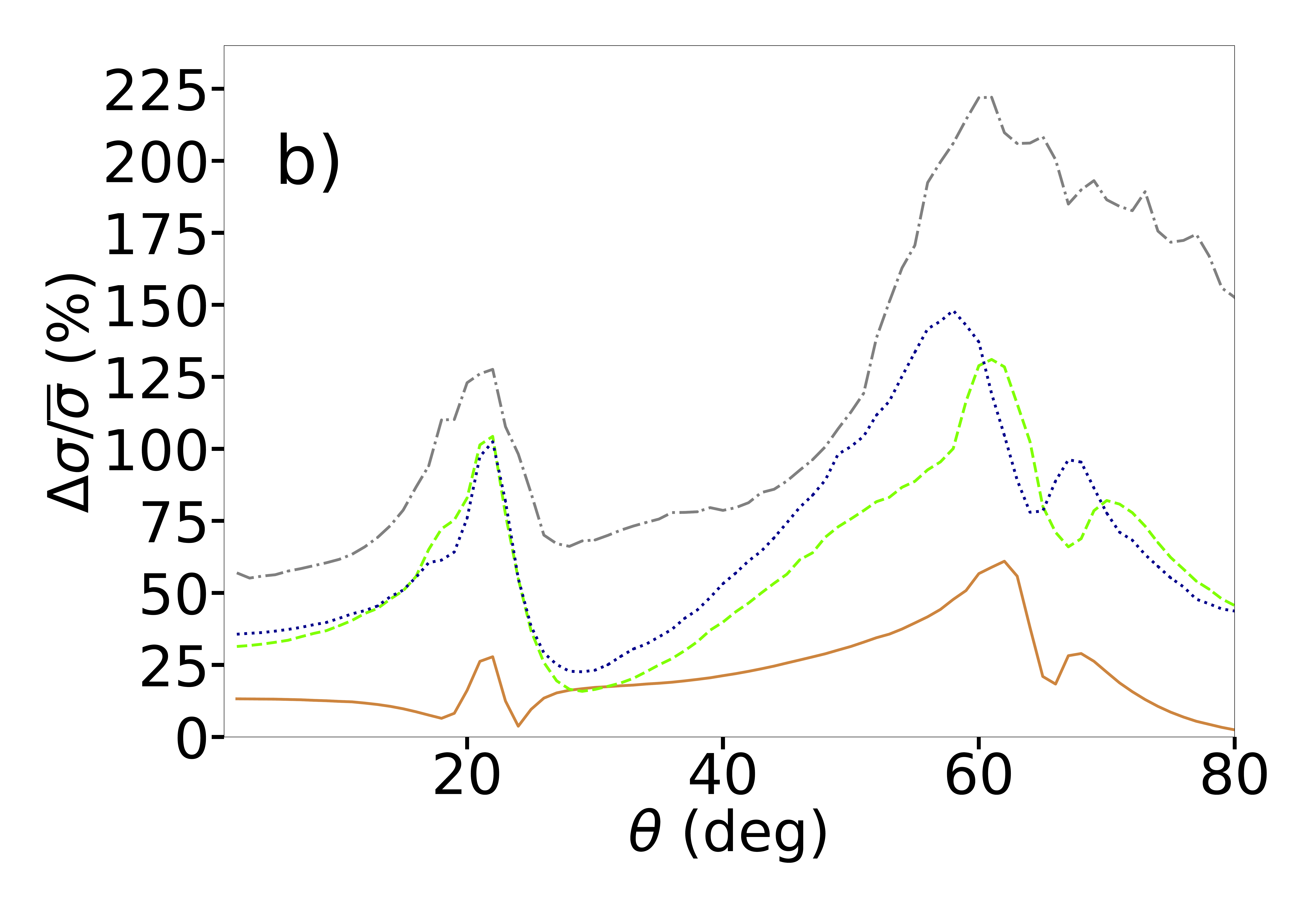}
    \includegraphics[width=8.5cm]{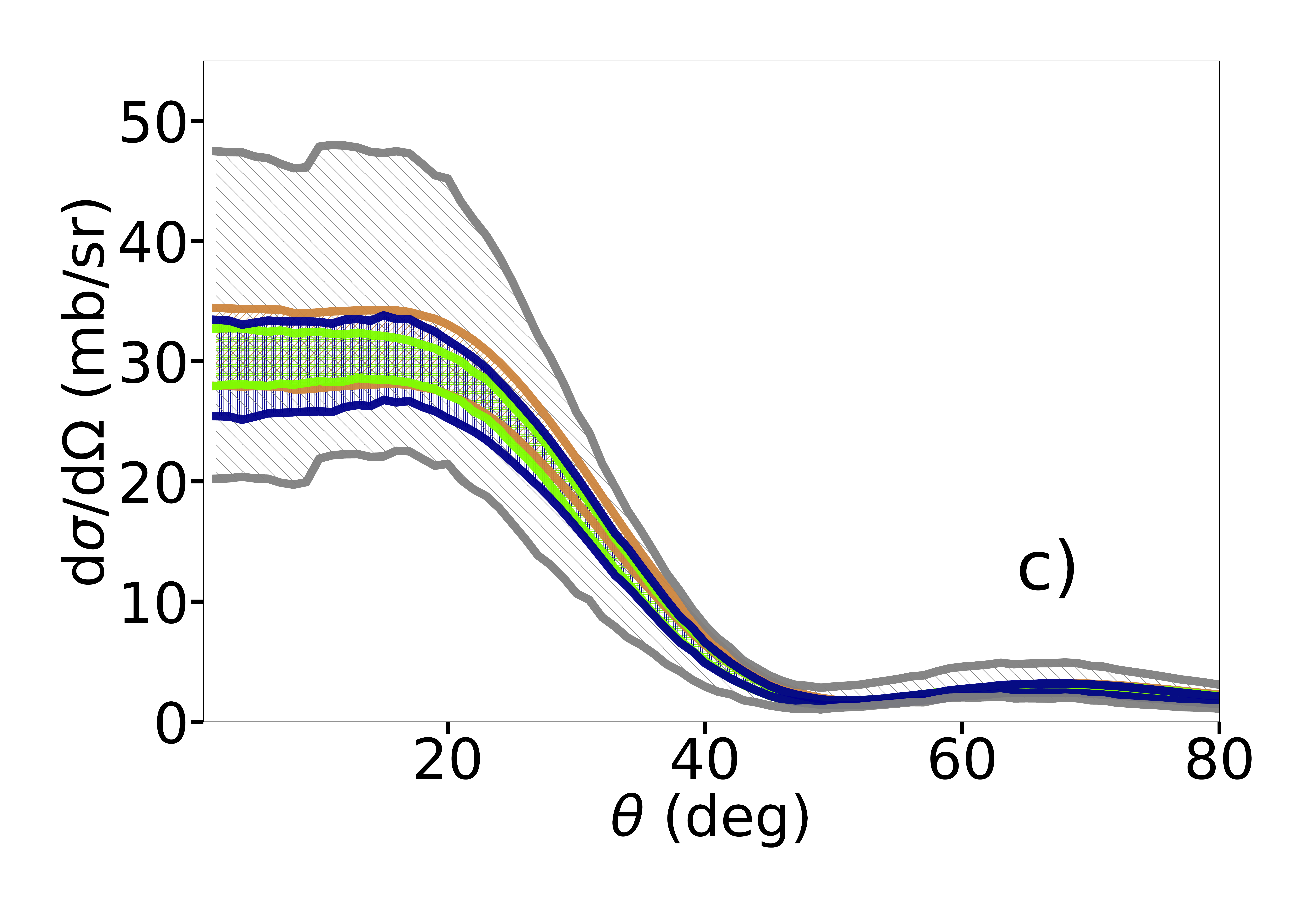}
    \includegraphics[width=8.5cm]{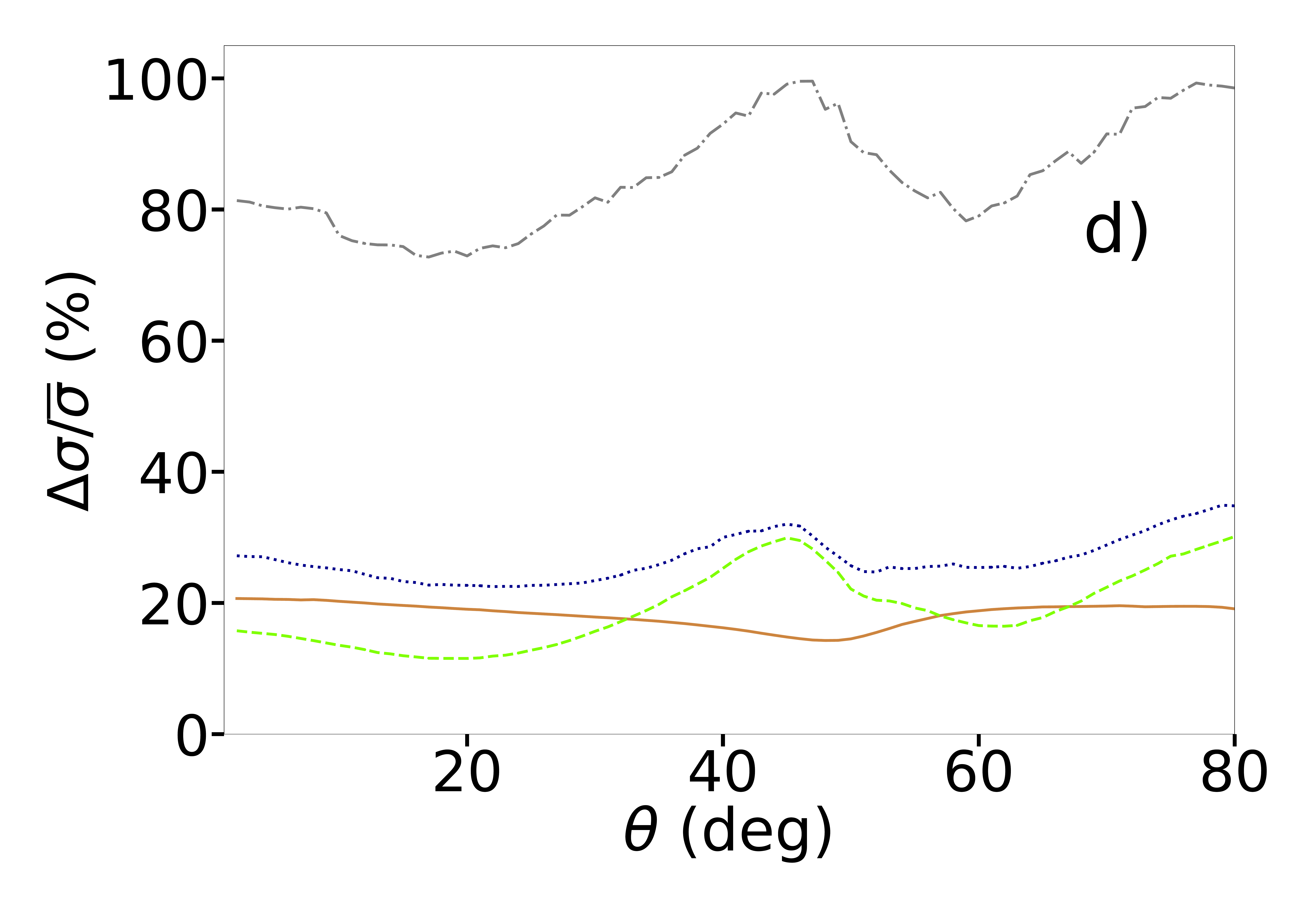}
    \includegraphics[width=8.5cm]{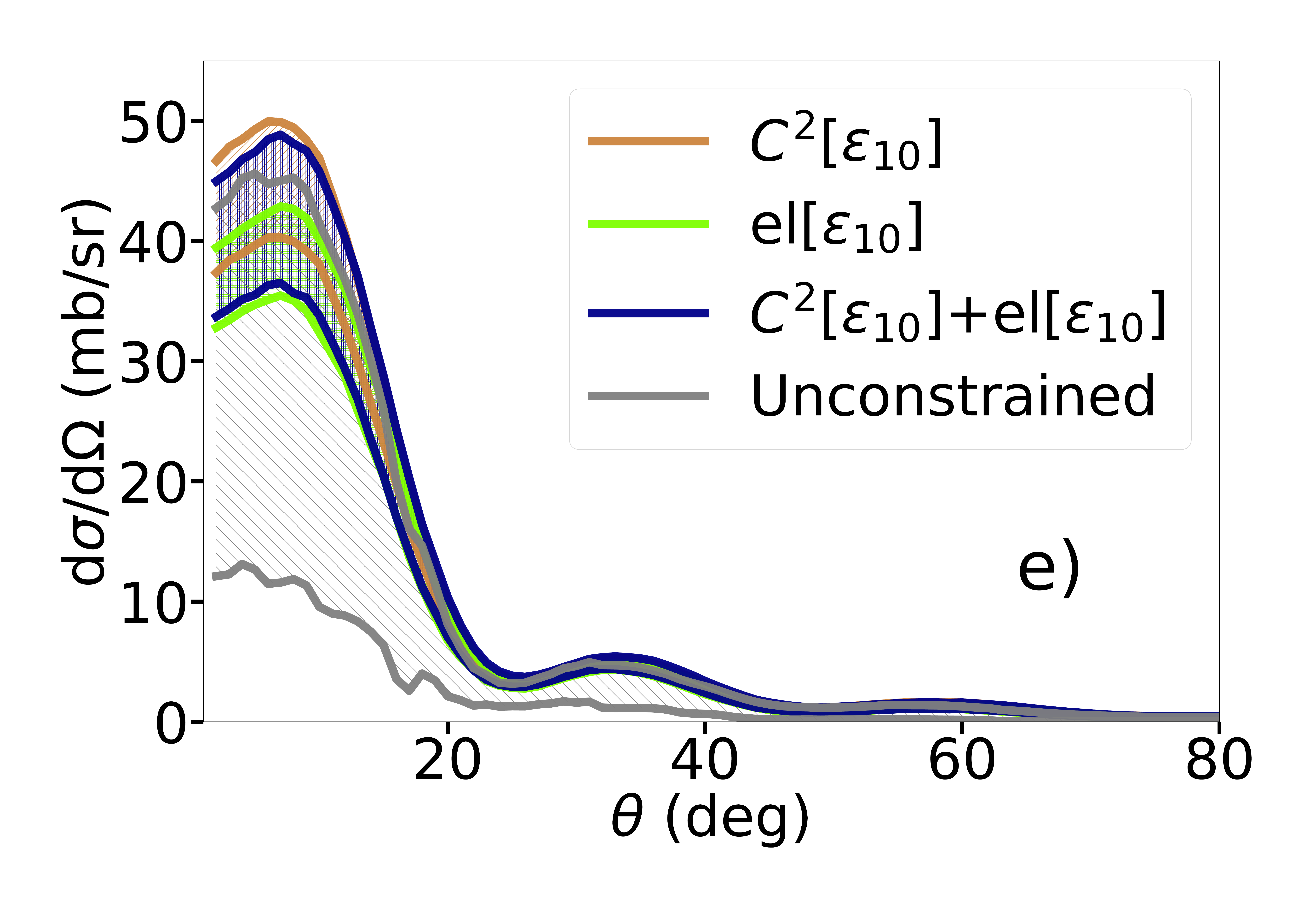}
    \includegraphics[width=8.5cm]{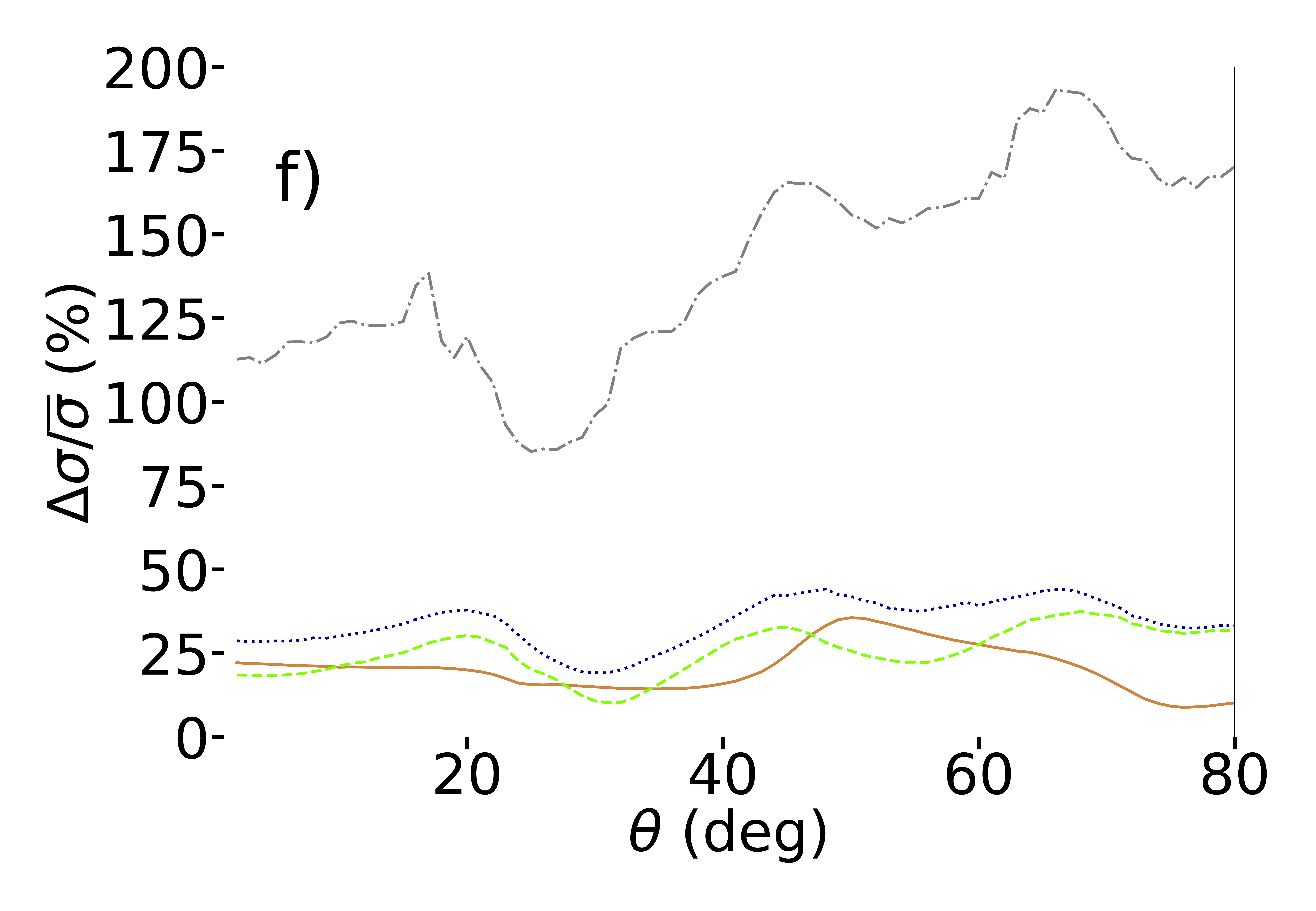}
    \caption{A comparison of results using only ANC ($C^2$) data in brown, only elastic data (el) in bright green and both types of data simultaneously ($C^2+\text{el}$), all at 10$\%$ error ($\varepsilon_{10}$) in blue, and unconstrained in gray; a) and b) $^{14}$C(d,p) at 17 MeV 68$\%$ confidence intervals and percentage uncertainty plot; c) and d) $^{16}$O(d,p) at 15 MeV 68$\%$ confidence intervals and percentage uncertainty plot; and e) and f) $^{48}$Ca(d,p) at 24 MeV 68$\%$ confidence intervals and percentage uncertainty plot}
    \label{fig:datatype_comp}
\end{figure*}
This analysis shows that constraining the bound state parameters is equally as important as constraining the nucleon-target interactions determining the incoming and outgoing channels. One might expect that the uncertainties coming from these different interactions be independent of one another. Our results are consistent with this expectation:  the quadrature sum of the error obtained constraining only the bound state interaction with the ANC and the error obtained constraining only the nucleon-target optical potentials with elastic scattering is equal to the square of the error of constraining both simultaneously, as expected for independent errors. 

Constraining only the bound state parameters or only the optical model parameters provides an inaccurate account of the uncertainties. Note that the brown and bright green bands in Fig. 3, which correspond to either constraining the bound state interaction only or the optical model parameters only, while fixing the other parameters, are not realistic predictions for uncertainties. These calculations were done to quantify the effects of these two different sources of uncertainty. However, they rely on  the unphysical assumption that the fixed parameters are known exactly (with zero error). To assess the uncertainty when no constraint is imposed on the parameters, we take $100$\% error on both ANC and elastic angular distributions, a way of representing minimal knowledge (shown in grey).

In Fig. 4, we compare the results for the physical situation when only one aspect is constrained, namely just the bound state interaction with the ANC (red) or the optical potentials with elastic scattering (green), and the other parameters (so-called unconstrained parameters) have posteriors corresponding to an experimental error of $100$\%. For comparison, we also include in Fig. 4, the same unconstrained result of Fig. 3 (gray), and the results obtained when both, bound state and scattering potentials, are constrained with precise data $\varepsilon_{10}$ (blue).
As before, the $68\%$ confidence intervals for the predicted transfer cross section are shown on the left panels: a) $^{14}$C$(d,p)^{15}$C(g.s.), c) $^{16}$O$(d,p)^{17}$O(g.s.), and e) $^{48}$Ca$(d,p)^{49}$Ca(g.s.). The corresponding percent uncertainties as a function of angle are shown on the right panels b), d), and f). 

Comparing the blue and gray bands in Fig. 4 gives a good estimate of the incredible improvement obtained when constraining the interactions using quality data. Comparing the green and red bands to the blue band, it is evident that there is only partial gain obtained by just constraining part of the interactions. For all cases, the uncertainties can be brought down to $20-30$\% if both ANC and elastic scattering constraints are included. 
It is also clear, comparing the red and the grey lines in  Fig. 4b, that the ANC does not offer an important constraint for the reaction on $^{14}$C, whereas it produces a large reduction of the uncertainty in the forward angle cross sections for the reactions on $^{16}$O and $^{48}$Ca  (red and grey lines in Fig. 4d and 4f). Through calculations not shown here, we determined that the forward-angle cross section for $^{14}$C(d,p) at 17 MeV  is not sensitive to the wavefunction at small distances, contrary to the (d,p) reactions on $^{16}$O and $^{48}$Ca here considered. 
We find that for the (d,p) reaction populating the halo state in $^{15}$C, it is the constraint on the elastic scattering that offers the best uncertainty reduction. 

As before, we can quantify the uncertainties by looking at the percent error at the peak of the transfer angular distribution (shown in Table IV).  

\begin{figure*}[t]
    \includegraphics[width=8.5cm]{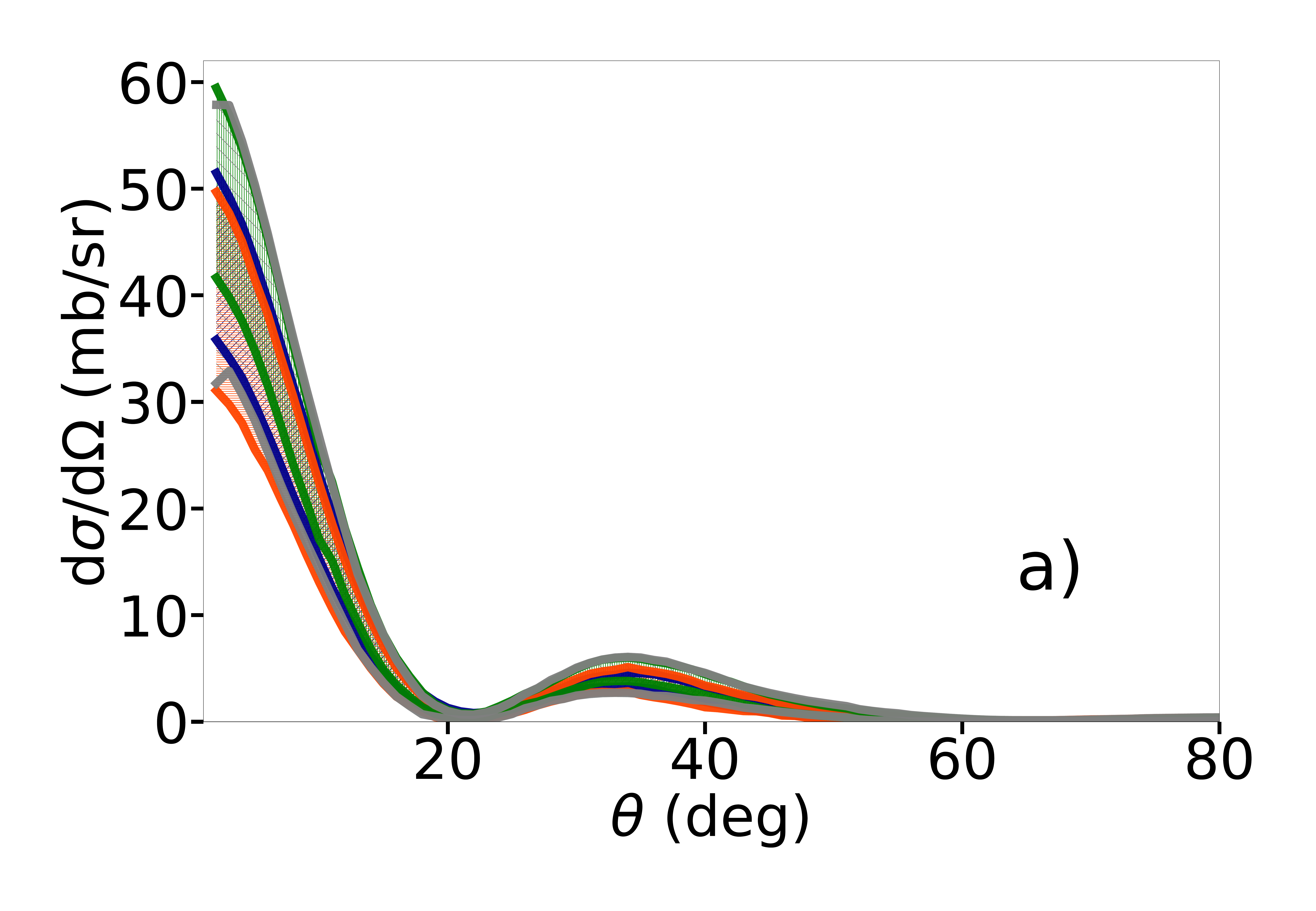}
    \includegraphics[width=8.5cm]{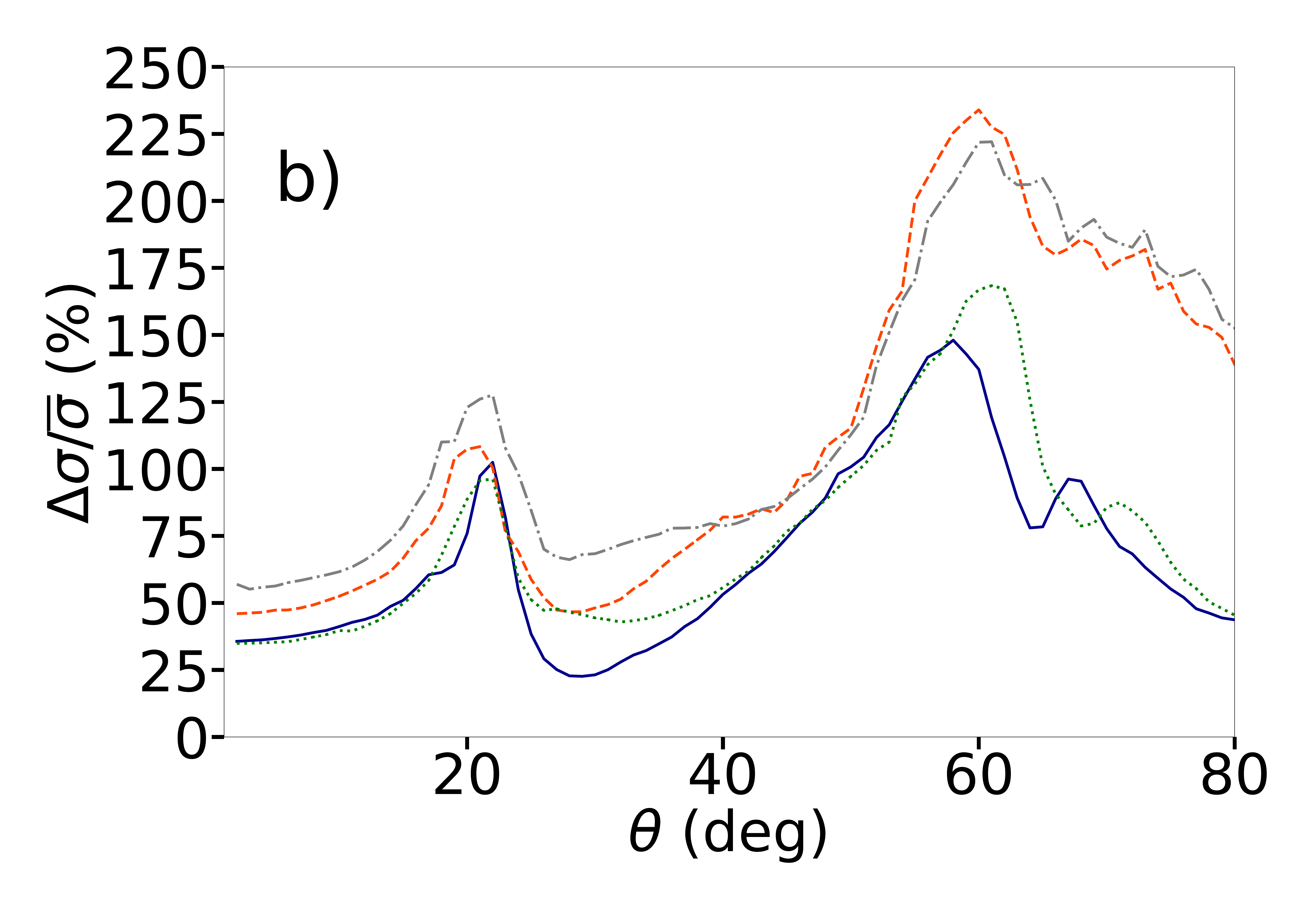}
    \includegraphics[width=8.5cm]{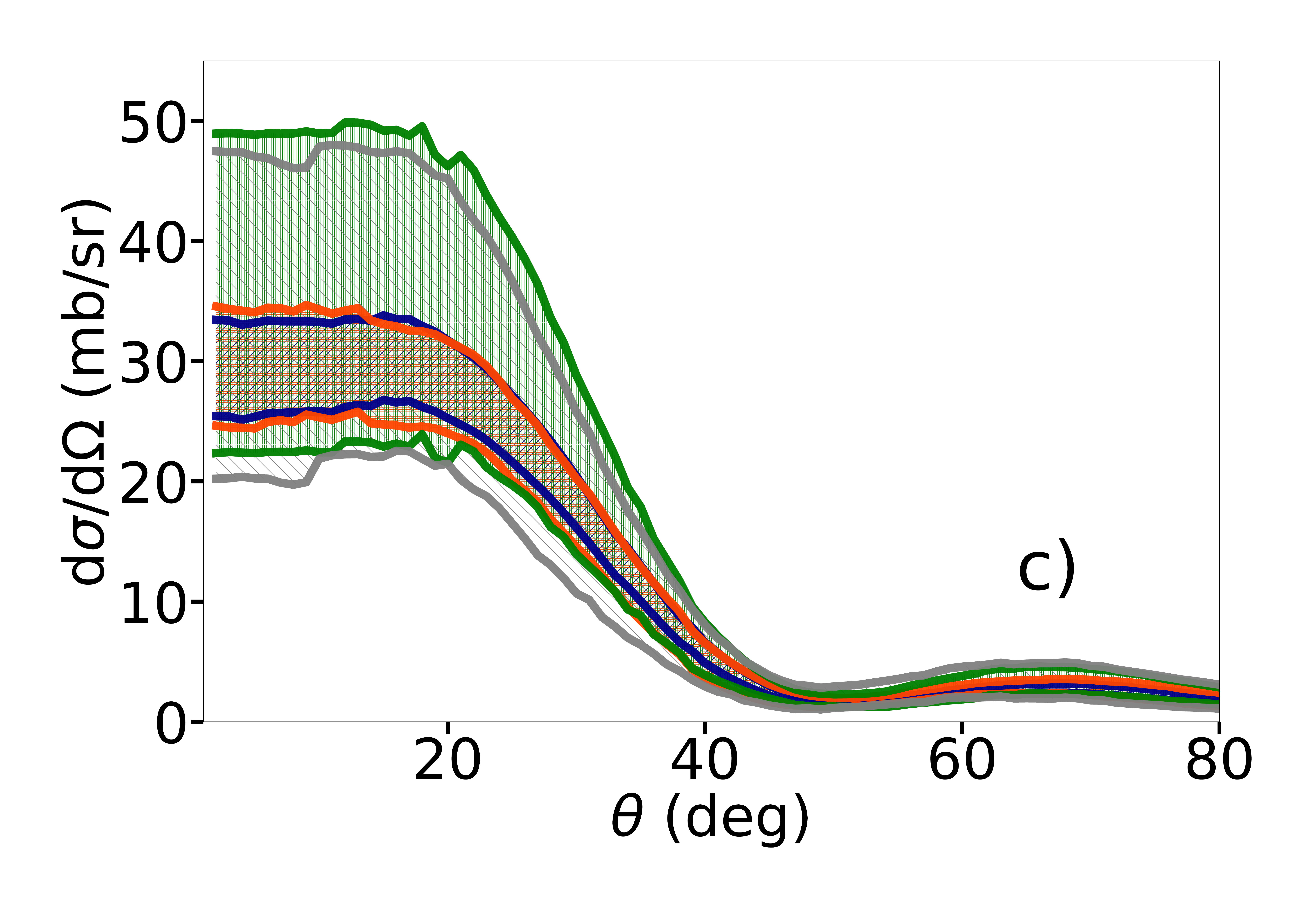}
    \includegraphics[width=8.5cm]{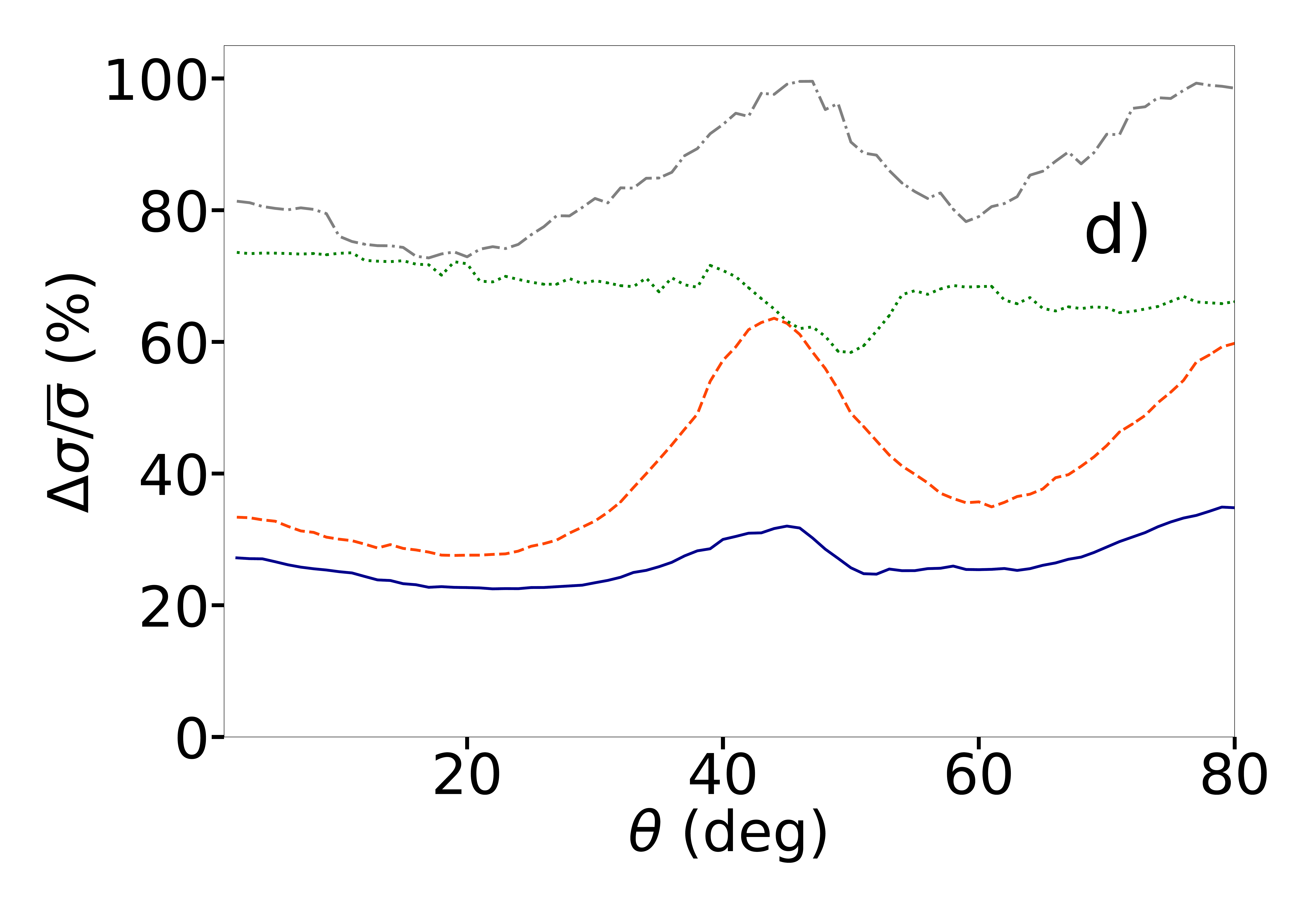}
    \includegraphics[width=8.5cm]{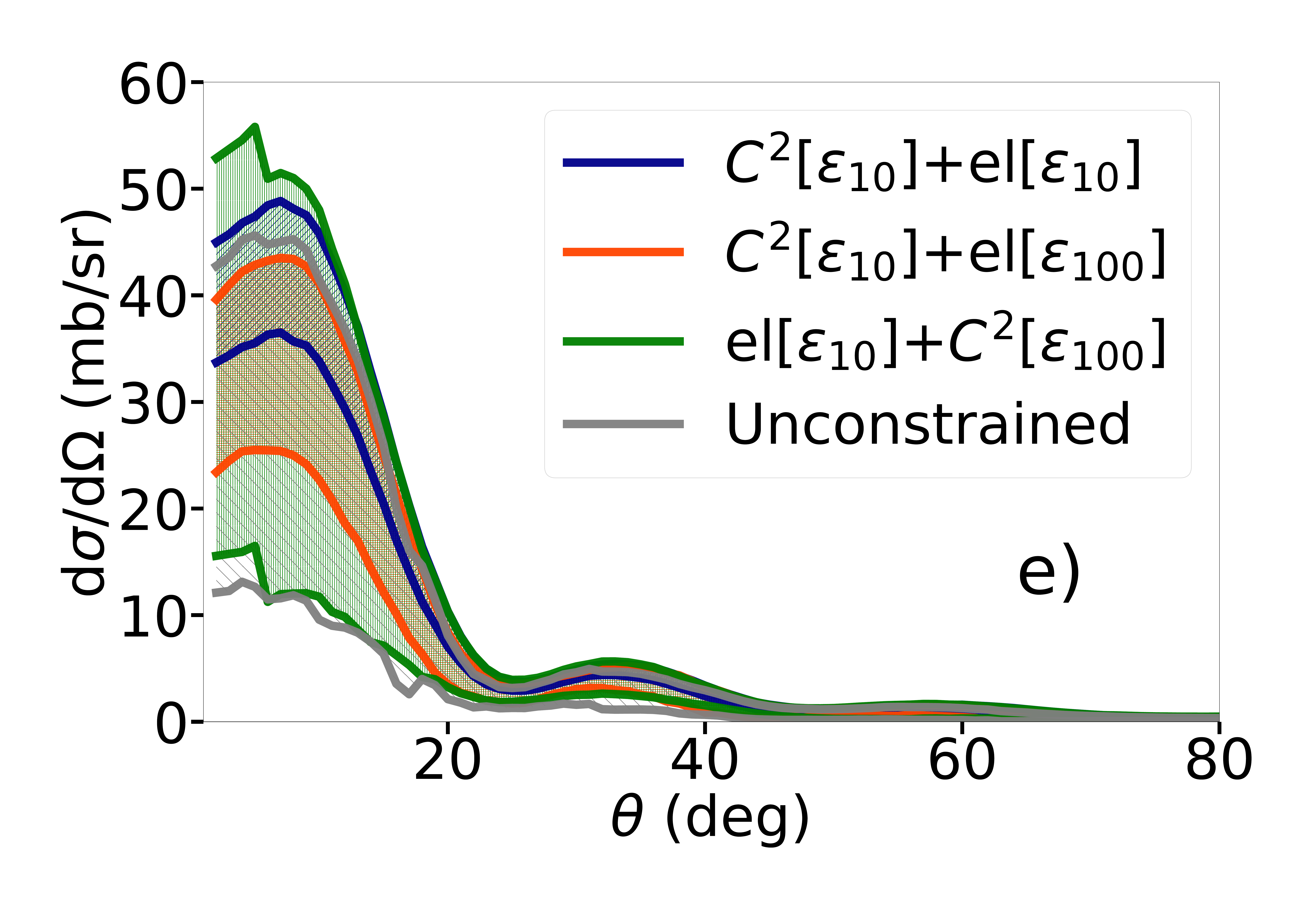}
    \includegraphics[width=8.5cm]{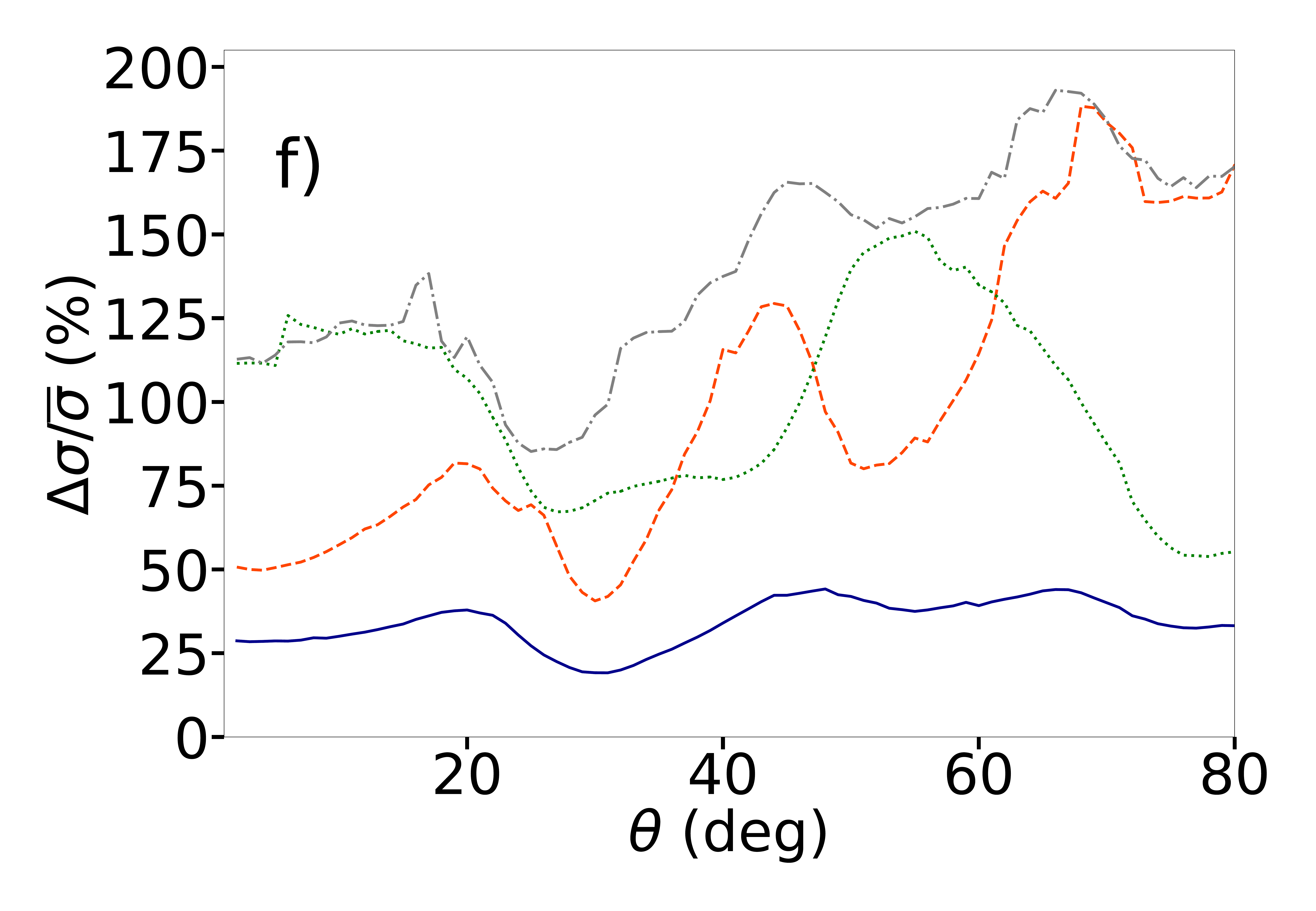}
    \caption{A comparison of results using different combinations of ANC ($C^2$) and elastic (el) constrains, $10\%$ error on both the ANC and elastic data in blue, 10$\%$ error on the ANC and $100\%$ error on the elastic in red, 10$\%$ error on the elastic and $100\%$ error on the ANC  in green, and unconstrained ANC and elastic data ($100\%$ error on both) in gray; a) and b) $^{14}$C(d,p) at 17 MeV 68$\%$ confidence intervals and percentage uncertainty plot; c) and d) $^{16}$O(d,p) at 15 MeV 68$\%$ confidence intervals and percentage uncertainty plot; and e) and f) $^{48}$Ca(d,p) at 24 MeV 68$\%$ confidence intervals and percentage uncertainty plot.}
    \label{fig:constrain_comp}
\end{figure*}


\subsection{The effect of the experimental error on parametric uncertainties}

Finally, given the rapid advances in beam intensities and detector systems, it is interesting to consider the improvement that can be obtained with high-precision experimental measurements.
 In this section we show the effects of decreasing the error in the data from $10$\% to $5$\%.
Fig. 5 is similar to Fig. 4:  the left panels depict the $68\%$ confidence intervals of the predicted transfer cross sections, a) $^{14}$C$(d,p)^{15}$C(g.s.), c) $^{16}$O$(d,p)^{17}$O(g.s.), and e) $^{48}$Ca$(d,p)^{49}$Ca(g.s.) and the right panels are the corresponding angular percent errors. We plot the results when all of the parameters are constrained by data with $5\%$ errors (magenta) to be compared with the results obtained before, with $10\%$ error (blue).  For reference we also include the unconstrained case, corresponding to results obtained when  the data has $100\%$ error (grey). 

\begin{figure}[t]
    \includegraphics[width=4.25cm]{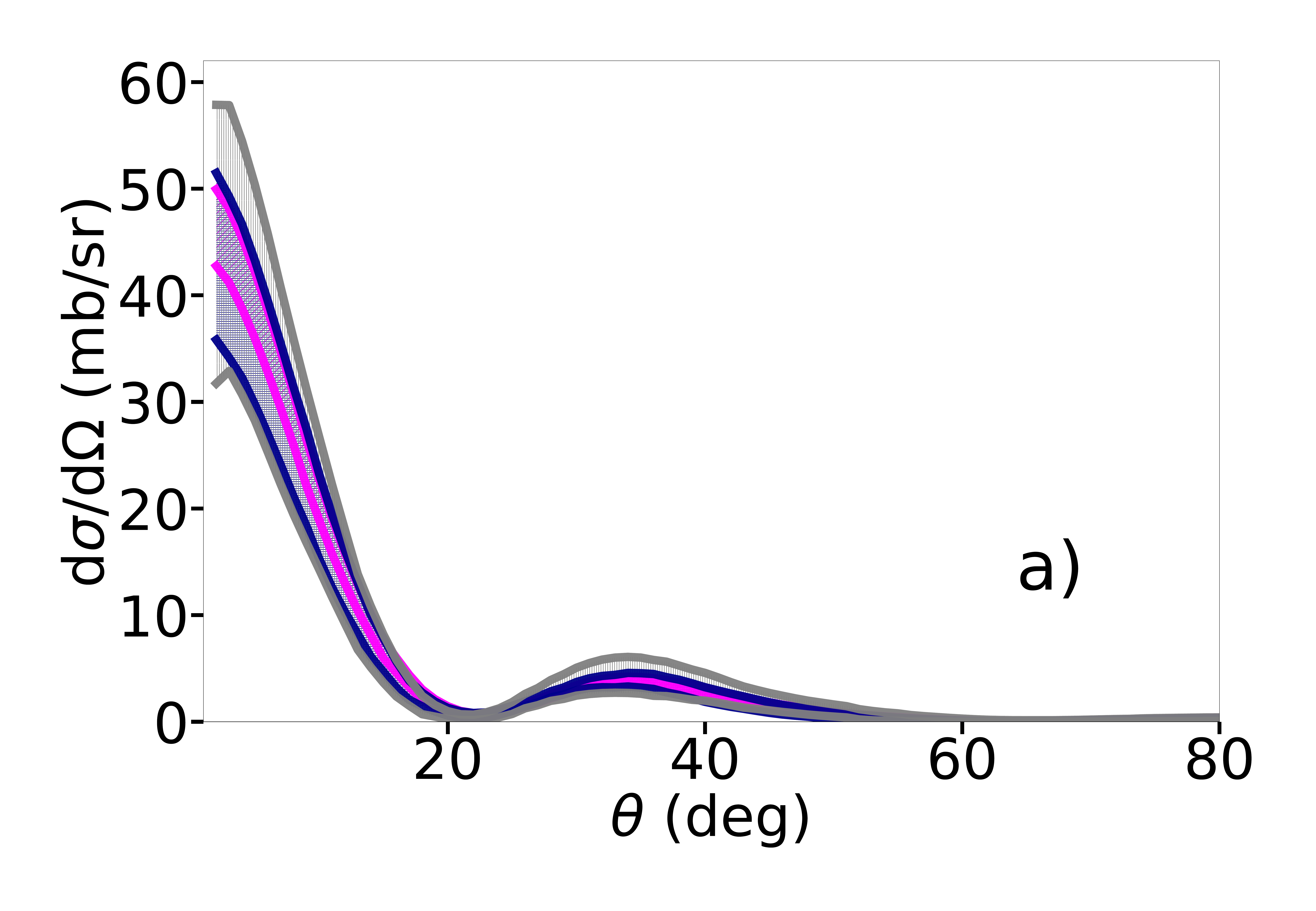}
    \includegraphics[width=4.25cm]{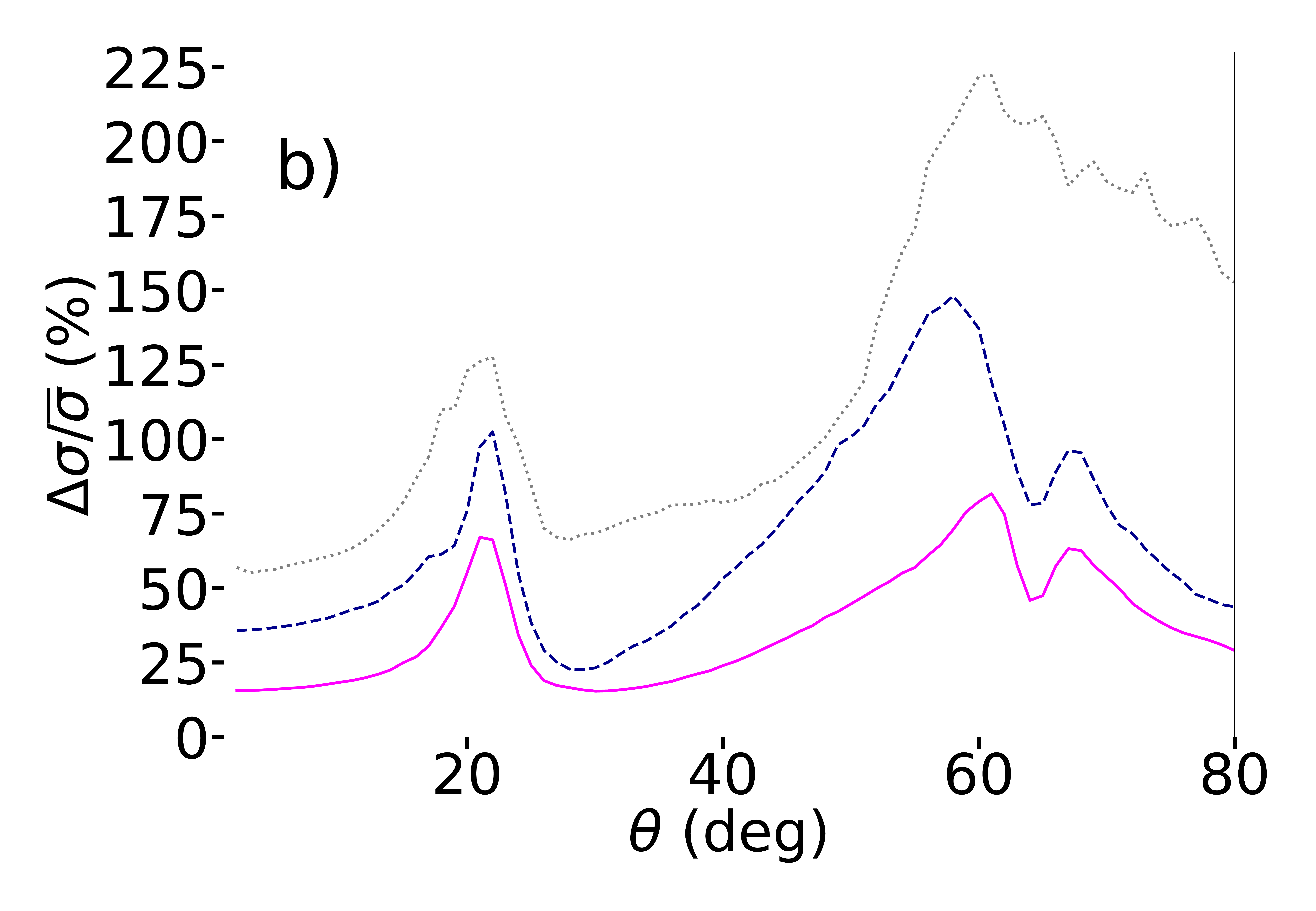}
    \includegraphics[width=4.25cm]{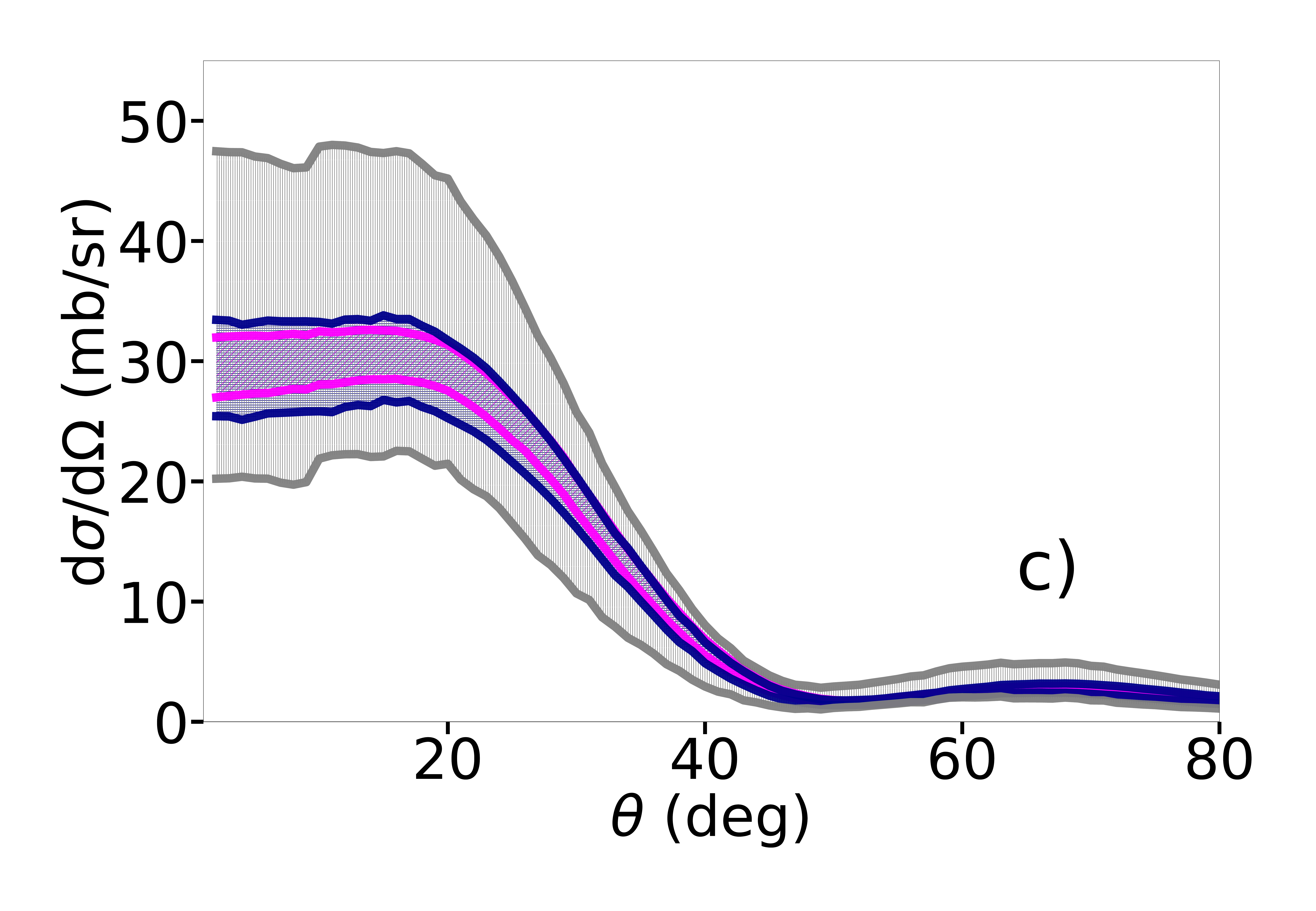}
    \includegraphics[width=4.25cm]{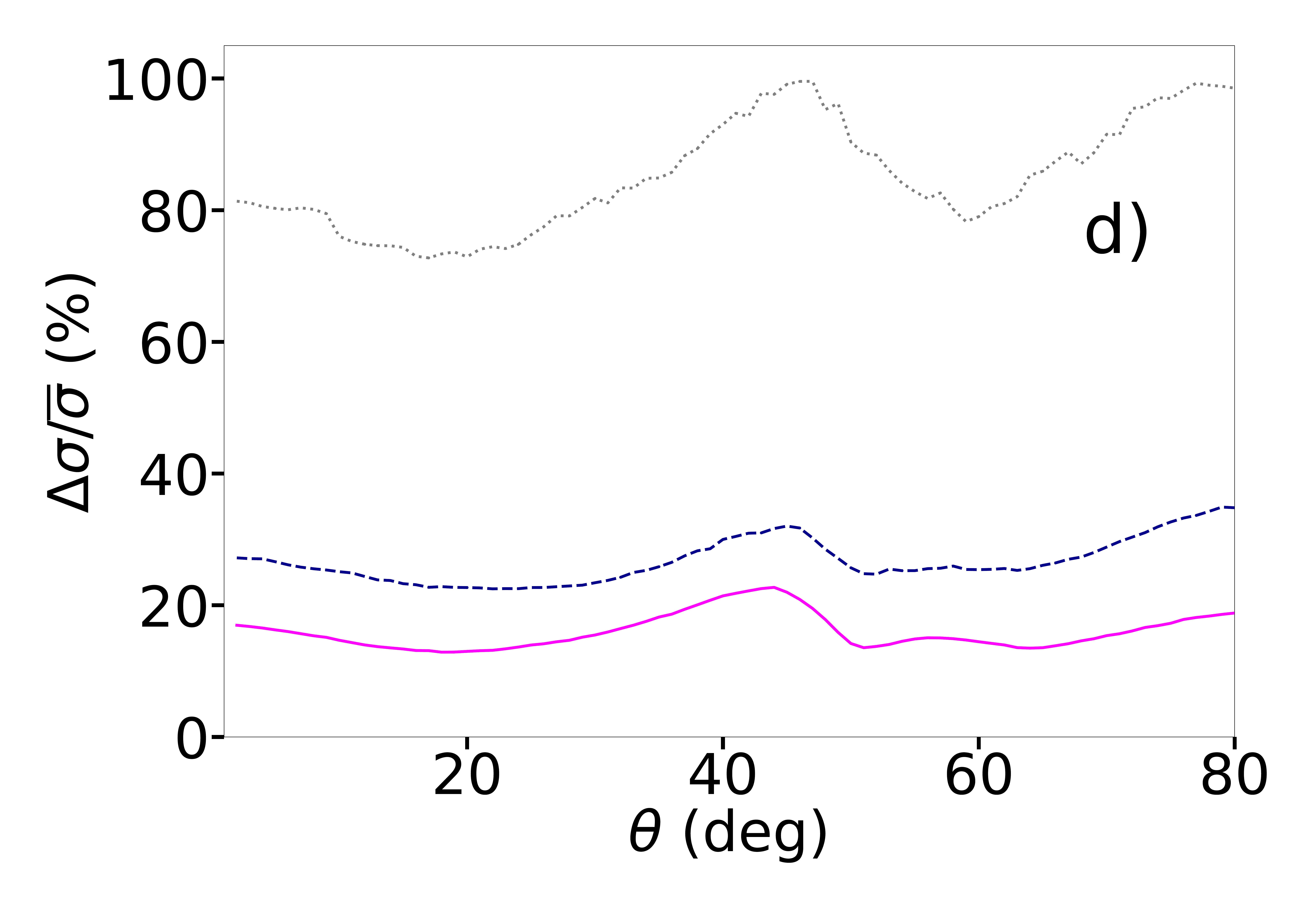}
    \includegraphics[width=4.25cm]{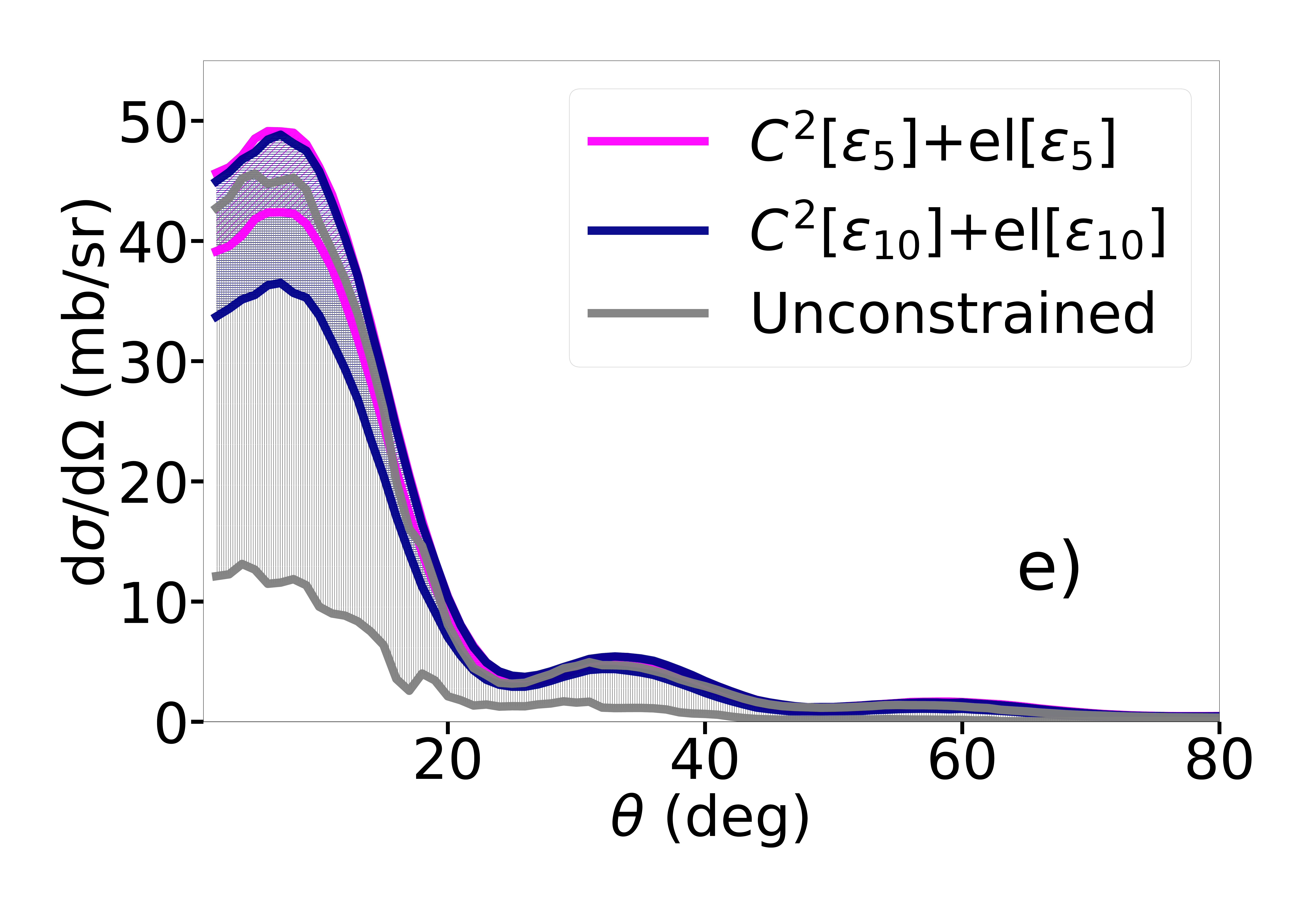}
    \includegraphics[width=4.25cm]{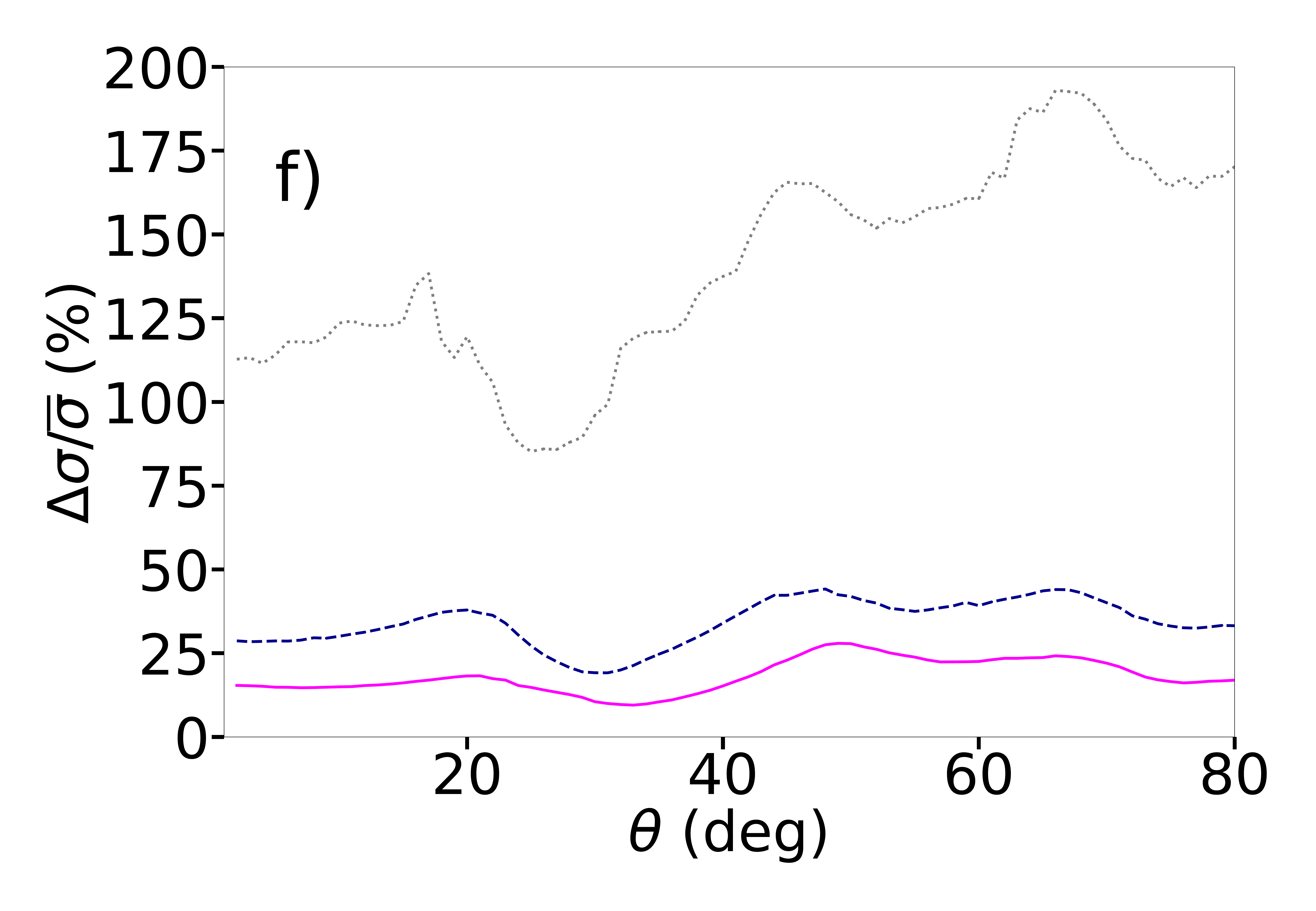}
    \caption{A comparison of results using different errors on both ANC ($C^2$) and elastic (el) data, $5\%$ error in magenta, $10\%$ error in blue and unconstrained date ($100\%$ error) in gray; a) and b) $^{14}$C(d,p) at 17 MeV 68$\%$ confidence intervals and percentage uncertainty plot; c) and d) $^{16}$O(d,p) at 15 MeV 68$\%$ confidence intervals and percentage uncertainty plot; and e) and f) $^{48}$Ca(d,p) at 24 MeV 68$\%$ confidence intervals and percentage uncertainty plot.}
    \label{fig:optimize_comp}
\end{figure}

We can observe from the percent errors in Fig. 5b), 5d), and 5f) that the decrease in experimental error unambiguously decreases the uncertainty in the predicted transfer cross section. This uncertainty is quantified in Table IV. 
Decreasing the experimental error from $10\%$ to $5\%$, reduces the percent error of the predicted cross section by roughly a factor of 2.

\begin{table*}[htp]
 \centering
 \begin{tabular}{c c| c|c| c c}
       Reaction & E (MeV) & Data [error] & $(\Delta\sigma/\bar{\sigma})^{68}_{\text{peak}}(\%)$ & $(\Delta\sigma/\bar{\sigma})^{95}_{\text{peak}}(\%)$ \\
      \hline \hline  
      $^{14}$C(d,p)$^{15}$C(g.s.) & 17  & $C^{2}[ \varepsilon_{10} ]$ & 13.2 & 25.7 \\
       $^{14}$C(d,p)$^{15}$C(g.s.) & 17   & $el[ \varepsilon_{10} ]$ & 31.4 & 65.1 \\
       $^{14}$C(d,p)$^{15}$C(g.s.) & 17  & $C^{2}[ \varepsilon_{5} ] + el[ \varepsilon_{5} ]$ & 15.6 & 30.1 \\
       $^{14}$C(d,p)$^{15}$C(g.s.) & 17   & $C^{2}[ \varepsilon_{10} ] + el[ \varepsilon_{10} ]$ & 35.7 & 73.0  \\
       $^{14}$C(d,p)$^{15}$C(g.s.) & 17   & $C^{2}[ \varepsilon_{10} ] + el[ \varepsilon_{100} ]$ & 46.0 & 103.3  \\
       $^{14}$C(d,p)$^{15}$C(g.s.) & 17  & $el[ \varepsilon_{10} ] + C^{2}[ \varepsilon_{100} ]$ & 34.9 & 77.9 \\
       $^{14}$C(d,p)$^{15}$C(g.s.) & 17  & $C^{2}[ \varepsilon_{100} ] + el[ \varepsilon_{100} ]$ & 56.9 &  140.4 \\
     \hline  
      $^{16}$O(d,p)$^{17}$O(g.s.) & 15  & $C^{2}[ \varepsilon_{10} ]$ & 20.7 & 39.5 \\
       $^{16}$O(d,p)$^{17}$O(g.s.) & 15  & $el[ \varepsilon_{10} ]$ & 15.5 & 34.3 \\
       $^{16}$O(d,p)$^{17}$O(g.s.)& 15   & $C^{2}[ \varepsilon_{5} ] + el[ \varepsilon_{5} ]$ & 13.5 & 33.4 \\
       $^{16}$O(d,p)$^{17}$O(g.s.) & 15  & $C^{2}[ \varepsilon_{10} ] + el[ \varepsilon_{10} ]$ & 23.3 & 54.9 \\
       $^{16}$O(d,p)$^{17}$O(g.s.) & 15  & $C^{2}[ \varepsilon_{10} ] + el[ \varepsilon_{100} ]$ & 30.3 & 71.4 \\
       $^{16}$O(d,p)$^{17}$O(g.s.) & 15  & $el[ \varepsilon_{10} ] + C^{2}[ \varepsilon_{100} ]$ & 72.4 & 151.8\\
       $^{16}$O(d,p)$^{17}$O(g.s.) & 15  & $C^{2}[ \varepsilon_{100} ] + el[ \varepsilon_{100} ]$ & 75.2 & 172.4 \\
       \hline $^{48}$Ca(d,p)$^{49}$Ca(g.s.) & 24 & $C^{2}[ \varepsilon_{10} ]$ & 21.4 & 42.4 \\
       $^{48}$Ca(d,p)$^{49}$Ca(g.s.) & 24  & $el[ \varepsilon_{10} ]$ & 18.9 & 37.6 \\
       $^{48}$Ca(d,p)$^{49}$Ca(g.s.) & 24  & $C^{2}[ \varepsilon_{5} ] + el[ \varepsilon_{5} ]$ & 14.8 & 29.4 \\
       $^{48}$Ca(d,p)$^{49}$Ca(g.s.) & 24  & $C^{2}[ \varepsilon_{10} ] + el[ \varepsilon_{10} ]$ & 28.9 & 56.0\\
       $^{48}$Ca(d,p)$^{49}$Ca(g.s.) & 24  & $C^{2}[ \varepsilon_{10} ] + el[ \varepsilon_{100} ]$ & 52.2 & 126.4  \\
       $^{48}$Ca(d,p)$^{49}$Ca(g.s.) & 24  & $el[ \varepsilon_{10} ] + C^{2}[ \varepsilon_{100} ] $ & 110.8 & 178.9 \\
       $^{48}$Ca(d,p)$^{49}$Ca(g.s.) & 24  & $C^{2}[ \varepsilon_{100} ] + el[ \varepsilon_{100} ]$ & 112.8 & 185.3 \\
 \end{tabular}
 \caption{Summary of the effect of propagating different combinations of the ANC ($C^2$) and elastic (el) uncertainties to the peak of the transfer angular distribution. For each reaction and energy considered (columns 1 and 2), and data included in the calculations along with a corresponding error (column 3), the percent error at the peak of the angular distribution of 68$\%$ confidence intervals (column 4) and 95$\%$ confidence intervals (column 5) are included.}
\end{table*}

\section{Conclusions}
\label{Conclusion}

In this study we present the first complete quantification of parametric uncertainties in (d,p) transfer cross sections. We extend previous work focused on the quantification of uncertainties from the optical potentials, to include the uncertainties associated with the final bound state. While the optical potential parameters are constrained through elastic scattering mock data, the bound state is constrained with the asymptotic normalization coefficient extracted from an independent measurement. This choice is based on previous work that indicated the usefulness of the ANC in reducing the ambiguity of the bound state overlap function.

As in previous studies, we use a Bayesian MCMC framework to determine parameter posterior distributions, and propagate these to the transfer cross sections, generating the $68\%$ and $95$\% confidence intervals for the angular distributions. We consider three reactions: $^{14}$C(d,p)$^{15}$C(g.s.) at $E_d=17$ MeV; $^{16}$O(d,p)$^{17}$O(g.s.) at $E_d=15$ MeV; and $^{48}$Ca(d,p)$^{49}$Ca(g.s.) at $E_d=24$ MeV. These reactions include a wide range of separation energies and angular momentum of the final state. Of the three, the reaction on $^{14}$C is unique because it populates an $s-$wave halo state that is loosely bound. 

We compare results using a standard $10\%$ error on the data (an error achievable for many current experiments), versus a $100\%$ error on the data, the later representing minimal information from experiment. Our results demonstrate conclusively that introducing the additional constraint on the bound state parameters through the ANC, on top of the constraints on the optical potential parameters through the elastic scattering data, reduces the uncertainty on the transfer cross section. This reduction is more noticeable for the reactions more sensitive to the interior, because it is for those reactions that the ambiguities associated with the bound state mean field are the largest.

We also show how misleading the uncertainty quantification can be when one ignores the uncertainties associated with a set of interactions. Keeping the interactions fixed (assuming zero error) leads to erroneously small uncertainties that do not correspond to reality. The low limit for the percent width of the $68\%$ confidence interval on the transfer cross section when minimal information is available on both optical potentials and bound state interactions is of $\sim 100\%$. This number is greatly reduced by introducing constraints on the optical potential and the bound state interaction: $\sim 30\%$.

Finally, we consider the prospect of having high-precision experiments with $5\%$ error, given the continual advances in  accelerator and detector technologies. For such cases, the percent $1\sigma$ width of the transfer cross section angular distribution stays consistently around $15\%$ for all reactions considered, a factor of two lower than that obtained when the experimental data have a $10\%$ error.

This work relies on a specific reaction model, namely ADWA. Although it accounts for deuteron breakup, it simplifies the deuteron three-body wavefunction by making the adiabatic approximation. In addition, this study assumes only pairwise interactions. These simplifications may introduce model uncertainties which are not yet quantified. More work is needed to address model uncertainties. Given the computational cost of exact three-body calculations, a necessary step to proceed with the quantification of model uncertainties is the development of fast and reliable emulators, along the lines of what was done in \cite{surer2022}.

\bibliography{uq}

\end{document}